\documentclass[onecolumn]{aastex631}

\usepackage{amsmath}
\usepackage{booktabs}

\received{June 21, 2024}
\revised{September 18, 2024}
\accepted{October 22, 2024}

\begin{document}

\title{Telescope-to-Fireball Characterization of Earth Impactor 2022 WJ1}

\correspondingauthor{Theodore Kareta}
\email{tkareta@lowell.edu}

\author[0000-0003-1008-7499]{Theodore Kareta}
\affiliation{Lowell Observatory,
1400 W. Mars Hill Road,
Flagstaff, AZ 86001, USA}

\author[0000-0003-4166-8704]{Denis Vida}
\affiliation{Department of Physics and Astronomy, University of Western Ontario, London, Ontario, N6A 3K7, Canada}
\affiliation{Western Institute for Earth and Space Exploration, University of Western Ontario, London, Ontario, N6A 5B7, Canada}

\author[0000-0001-7895-8209]{Marco Micheli}
\affiliation{ESA PDO NEO Coordination Centre, Planetary Defence Office, Largo Galileo Galilei, 1, 00044 Frascati (RM), Italy}

\author[0000-0001-6765-6336]{Nicholas Moskovitz}
\affiliation{Lowell Observatory,
1400 W. Mars Hill Road,
Flagstaff, AZ 86001, USA}

\author[0000-0002-1914-5352]{Paul Wiegert}
\affiliation{Department of Physics and Astronomy, University of Western Ontario, London, Ontario, N6A 3K7, Canada}
\affiliation{Western Institute for Earth and Space Exploration, University of Western Ontario, London, Ontario, N6A 5B7, Canada}

\author[0000-0001-6130-7039]{Peter G. Brown}
\affiliation{Department of Physics and Astronomy, University of Western Ontario, London, Ontario, N6A 3K7, Canada}
\affiliation{Western Institute for Earth and Space Exploration, University of Western Ontario, London, Ontario, N6A 5B7, Canada}

\author[0000-0003-3030-7524]{Phil J. A. McCausland}
\affiliation{Department of Earth Sciences, University of Western Ontario, London, Ontario, N6A 5B7, Canada}
\affiliation{Western Institute for Earth and Space Exploration, University of Western Ontario, London, Ontario, N6A 5B7, Canada}

\author[0000-0001-9226-1870]{Hadrien A. R. Devillepoix}
\affiliation{School of Earth and Planetary Sciences, Curtin University, Perth WA 6845, Australia}
\affiliation{International Centre for Radio Astronomy Research, Curtin University, Perth WA 6845, Australia}

\author{Barbara Malečić}
\affiliation{University of Zagreb, Faculty of Science, Department of Geophysics, Horvatovac 95, 10000 Zagreb, Croatia}

\author{Maja Telišman Prtenjak}
\affiliation{University of Zagreb, Faculty of Science, Department of Geophysics, Horvatovac 95, 10000 Zagreb, Croatia}

\author{Damir \v{S}egon}
\affiliation{Astronomical Society Istra Pula, Park Monte Zaro 2, HR-52100 Pula, Croatia}
\affiliation{Vi\v{s}njan Science and Education Center, Istarska 5, HR-51463 Vi\v{s}njan, Croatia}

\author{Benjamin Shafransky}
\affiliation{Lowell Observatory,
1400 W. Mars Hill Road,
Flagstaff, AZ 86001, USA}

\author[0000-0003-0774-884X]{Davide Farnocchia}
\affiliation{Jet Propulsion Laboratory, California Institute of Technology, 4800 Oak Grove Dr, Pasadena, CA 91109, USA}
 
\begin{abstract}
Comparing how an asteroid appears in space to its ablation behaviour during atmospheric passage and finally to the properties of associated meteorites represents the ultimate probe of small near-Earth objects. We present observations from the Lowell Discovery Telescope and from multiple meteor camera networks of 2022 WJ1, an Earth impactor which was disrupted over the North American Great Lakes on 19 November 2022. As far as we are aware, this is only the second time an Earth impactor has been specifically observed in multiple passbands prior to impact to characterize its composition. The orbits derived from telescopic observations submitted to the Minor Planet Center (MPC) and ground-based meteor cameras result in impact trajectories that agree to within 40 meters, but no meteorites have been found as of yet. The telescopic observations suggest a silicate-rich surface, and thus a moderate-to-high albedo, which results in an estimated size for the object of just $D=40-60$ cm. Modeling the fragmentation of 2022 WJ1 during its fireball phase also suggests an approximate half-meter original size for the object as well as an ordinary chondrite-like strength. These two lines of evidence both support that 2022 WJ1 was likely an S-type chondritic object and the smallest asteroid compositionally characterized in space. We discuss how best to combine telescopic and meteor camera datasets, how well these techniques agree, and what can be learned from studies of ultra-small asteroids.
\end{abstract}


\section{Introduction} \label{sec:intro}
\subsection{Earth Impactors}
While the link between asteroids, meteors, and meteorites might be well understood in general, studying the same object in all three domains has been challenging. Just eight asteroids have been discovered in space prior to impacting the Earth as of mid-2024, usually with only a few hours notice. This means that while linking meteorites to meteors observed either visually or with automated camera systems is being done increasingly frequently, though many meteorites were observed `falls' well before the deployment of dedicated, modern video networks. However, characterizing these Earth impactors in space as one would a typical near-Earth asteroid (NEA) has required a set of rare circumstances -- and a significant amount of luck.

The benefits to being able to compare and contrast measurements between these three related fields is shown clearly by the discovery and characterization of 2008 TC$_3$ in space and its later recovery on the ground as the meteorite Almahatta Sitta \citep{2009Natur.458..485J}. Discovered about $\sim21$ hours prior to impact in the Sudanese desert, a reflectance spectrum obtained at the $4.2$ m William Herschel Telescope in the Canary Islands $\sim2.5$ hours prior to impact \citep{2009Natur.458..485J} found that the asteroid had a relatively flat or slightly blue-sloped reflectance spectrum at visible wavelengths, most similar to the dark and carbonaceous F-type asteroids \citep{1985Icar...61..355Z}. (F-type asteroids were absorbed into the broader B-type classification in later taxonomies like \citet{2002Icar..158..146B,2009Icar..202..160D}.) The meteorites on the ground were classified by \citet{2009Natur.458..485J} as ureilites, a kind of carbon-rich achondritic meteorite without a known parent body. The match between the B/F type asteroids and the ureilites is particularly surprising, as their silicate-rich nature led to the prediction of their association with the S-complex asteroids \citep{1993Metic..28..161G}. Later analyses by \citet{2010M&PS...45.1638B} showed that while most of the fragments of Almahatta Sitta found were ureilitic, a little less than half (17/40) were chondritic in origin -- a completely different kind of material formed under different conditions. \citet{2010M&PS...45.1618Z} also found significant heterogeneity among the samples, including down to the scale of samples of a few grams or less. While multiple meteorite falls happening in the same general area might be plausible, those authors estimated that at least seven separate and recent falls very close in proximity would be necessary to produce the variety of meteorite types seen if each fall only dropped a single kind of rock. Later work by \citet{2022M&PS...57.1641J} showed that the distribution of rock types on the ground required that the heterogeneity extended throughout the body of 2008 TC$_3$ itself. Several meteorite types originating from the same fall is also the likely explanation for Kaidun \citep{2003ChEG...63..185Z}, a meteorite which showed an even wider range of compositions in individual rocks. In other words, the asteroid 2008 TC$_3$ likely had considerable compositional heterogeneity despite being only a few meters across. When JAXA's Hayabusa-2 and NASA's OSIRIS REx arrived at the (larger) carbonaceous NEAs (162173) Ryugu and (101955) Bennu a decade later, the discovery of plentiful exogenous material at these objects \citep{2021NatAs...5...31D, 2021NatAs...5...39T} -- such as bright basalt rocks scattered about the surface of Bennu -- showed that heterogeneity might be a common feature of rubble-pile asteroids. The collisional history of these objects is written on their surfaces and encoded in the rocks that make it to the Earth's surface as meteorites.

In the case of 2008 TC$_3$ and later cases like 2018 LA (in which photometry was collected of the object prior to impact instead of spectroscopy, \citealt{2021M&PS...56..844J}), a comparison of telescopic observations of an asteroid against laboratory analyses of meteorites from that asteroid resulted in a much deeper understanding of that body's original properties -- not just its modern composition, but also insights into its structure and history. These kinds of interdisciplinary studies can thus shed light on numerous topics of common interest to the communities that study asteroids, meteors, and meteorites.

Similar to the case of 2008TC$_3$, the Earth impactor 2022 WJ1 (hereafter referred to as WJ1) was discovered by the Catalina Sky Survey at 04:53 UTC on 19 November 2022, approximately three hours prior to its impact which was predicted to occur in the vicinity of the Great Lakes region on the border of Canada and the United States. This was a fortuitous event, as not only was this enough time for a significant number of observers to report astrometry to refine WJ1's pre-impact orbit and likely area of impact, it was enough time for one large telescope to be activated to characterize the object -- namely, the $4.3$ m Lowell Discovery Telescope. Furthermore, WJ1's last seconds would be well captured by the University of Western Ontario's large meteor camera network. In other words, this was a clear opportunity to pursue exactly the kind of endeavour that was epitomized by the the case of 2008 TC$_3$ : a telescope-to-fireball characterization of an Earth impactor, with hopes for more if rocks were found on the ground.

In this paper, we present and discuss a combined analysis to understand this object as an asteroid in space and as a fireball over the Great Lakes. We first review the astrometric dataset, including the last telescopic detection of WJ1 $\sim40$ s prior to its entry into the shadow of the Earth, and discuss the object's pre-impact orbit and likely escape routes from the Main Belt. Second, we review the physical implications of the photometric observations obtained of WJ1 at the Lowell Discovery Telescope and discuss the object's likely composition, rotation state, and size as inferred from our measured broadband colors. Third, we present and synthesize observations of WJ1 as it ablated in the atmosphere, including constraints on its composition from material strength and where potential meteorites might be found. Lastly, we compare the conclusions drawn from the three lines of analysis to verify the efficacy of each kind of methodology, to constrain the properties of the smallest and most common asteroids, and to make recommendations for future Earth impactor observation schemes.

\section{Observations} \label{sec:obs}
Imaging observations of WJ1 were obtained with the Large Monolithic Imager (LMI, \citealt{2014SPIE.9147E..2NB}) on the 4.3 m Lowell Discovery Telescope (LDT) on 19 November 2022 in 3$\times$3 binning mode for an effective pixel scale of $0.36"$. While the discovery of WJ1 came just hours earlier, the observations presented and analyzed in this paper do not come from a Target-of-Opportunity interrupt at the telescope but simply due to good luck of having already been scheduled to observe mission-accessible near-Earth asteroids during that time slot. The LDT's exceptionally fast and stable tracking has made it an invaluable asset in characterizing small and fast moving NEAs, but at the time observations started the object had yet to have been given a designation and ephemerides were only available through JPL Scout and the MPC's NEO confirmation page, and thus differently formatted than the input the telescope control system (TCS) was designed for, namely a JPL Horizons query. 

Until the ephemeris files from Scout could be altered into a format that the TCS would accept (a task which would only be accomplished after WJ1 entered the shadow of the Earth and was no longer easily visible), the plan for the observations was to slew to the coordinates of where WJ1 would be in approximately sixty seconds and begin taking relatively short exposures as the object streaked through the frame of view. The object was sufficiently bright to be identified rapidly and unambiguously. As the object would leave the field of view of LMI, another filter would be selected, the telescope operator would slew the telescope, and the process would start again.  Several operational `lessons learned' from this case study are discussed in Section \ref{sec:disc}. In the following sub-sections, we detail the pre-impact orbit of the object as constrained by our astrometry and that from others as well as what can be learned about its rotation state, composition, and size from our multi-filter observations.

\subsection{Astrometric Coverage and Pre-Impact Orbit}

The available observational coverage of 2022~WJ1 in space spans about three hours, from the discovery observation at 04:53 UTC to just before its entry into Earth's shadow at 07:59 UTC. Seven observatories reported a total of 46 astrometric measurements to the Minor Planet Center (MPC) prior to the time of impact, already allowing for an accurate sub-km determination of the location of the pre-atmospheric impact point. Additional observations were reported by three additional stations during the following days and months, resulting in a complete astrometric dataset that now includes 51 astrometric measurements, from ten observatories located across the United States from Hawaii to Arizona and thus spanning a range of longitudes.

\begin{table}[h]
\caption{The Pre-Impact Orbit of 2022~WJ1, as determined from telescopic observations. All angular elements are Epoch J2000.}\label{tab:orbit}%
\resizebox{\textwidth}{!}{\begin{tabular}{@{}lll@{}}
\toprule
 Symbol & Parameter  & Telescopic-Derived Values\\
\midrule
$e$         & Eccentricity                             &   0.51314 $\pm$ 6.99e-5 \\
$a$         &  Semimajor axis (AU)                     &   1.90473 $\pm$ 2.58e-4 \\
$q$         &  Perihelion distance (AU)                &   0.92734 $\pm$ 7.67e-6 \\
$i$         & Inclination ($^{\circ}$)                  &   2.61764 $\pm$ 2.83e-4 \\
$\Omega$    & Longitude of ascending node ($^{\circ}$)  & 56.71527 $\pm$ 1.00e-5 \\
$\omega$    & Argument of perihelion ($^{\circ}$)       & 35.1056 $\pm$ 6.52e-4 \\
$M$         & Mean anomaly  ($^{\circ}$)                & 349.729 $\pm$ 2.20e-3 \\
            & Epoch (TDB)                              & 2459902.5             \\
\botrule
\end{tabular}}
\end{table}

A significant fraction of these observations have been reported to the MPC in the new ADES (Astrometry Data Exchange Standard) astrometric format, and most of them include formal uncertainties for the right ascension and declination measurements, computed by the observers. For those that do not include uncertainties, a standard conservative assumption of ±0.6" was used.
Unfortunately none of the available astrometric measurements include an estimate of instrumental timing uncertainties (e.g. due to offsets between the time stamps of individual exposures and the actual times that the instrument shutter opened). These are equally if not more important for an object that was moving at nearly one arcminute per second near the end of the observed arc. In the following we will assume a time uncertainty of 1 s for most stations, with the exception of those obtained with the University of Hawaii 2.2 m telescope and the Lowell Discovery Telescope, which are known to be accurate to ±0.1 s or better when corrected for known time biases. The time biases at these facilities have been confirmed by comparing measured on-sky positions of Global Navigation Satellite Systems (GNSS) to ephemeris predictions.

This dataset forms the basis for our orbit determination of 2022~WJ1 from pre-entry telescopic data as summarized in Table \ref{tab:orbit}. The LDT astrometry included in this analysis is obtained from the dataset presented in this work, and has been extracted by our team with proper trail-fitting procedures. It represents the latest known detection of the object before its entry into Earth's shadow, and provides significant information for the orbit determination process.

The best-fit orbit for WJ1 prior to impact is not an uncommon one for an Earth impacting NEA, with a perihelion just interior to the Earth ($q=0.927$ au), a semi-major axis in the inner Main Belt ($a=1.905$ au), and a low inclination ($i=2.62^{\circ}$). The object was inbound to perihelion when it impacted the Earth. The JPL Small Body Database quotes an absolute magnitude of $H_V=33.58\pm0.36$ for WJ1, which implies a likely diameter of around a meter for typical asteroidal albedos ($0.04 < p_V < 0.45$). If one assumes that the modeled distribution of NEA orbits from \citet{2018Icar..312..181G} can be extrapolated down to such small sizes (the model is only formally applicable to $17 < H_V < 25$), WJ1 has a $\sim82\%$ chance of having escaped the Main Belt via the $\nu_6$ resonance, a $9\%$ chance of having escaped from the $3:1$ resonance, and a $9\%$ chance of having originated in the Hungarias \citep[values retrieved from the astorb database,][]{2022A&C....4100661M}. All of these source regions are dominated by S-complex asteroids (see, e.g., \citealt{2014Natur.505..629D}); thus if the Granvik model is applicable to WJ1, a stony composition and moderate-albedo ($0.15 < p_V < 0.45$) seems likely.

\subsubsection{Visibility of WJ1 in Prior Epochs}
\begin{figure}[ht!]
\plotone{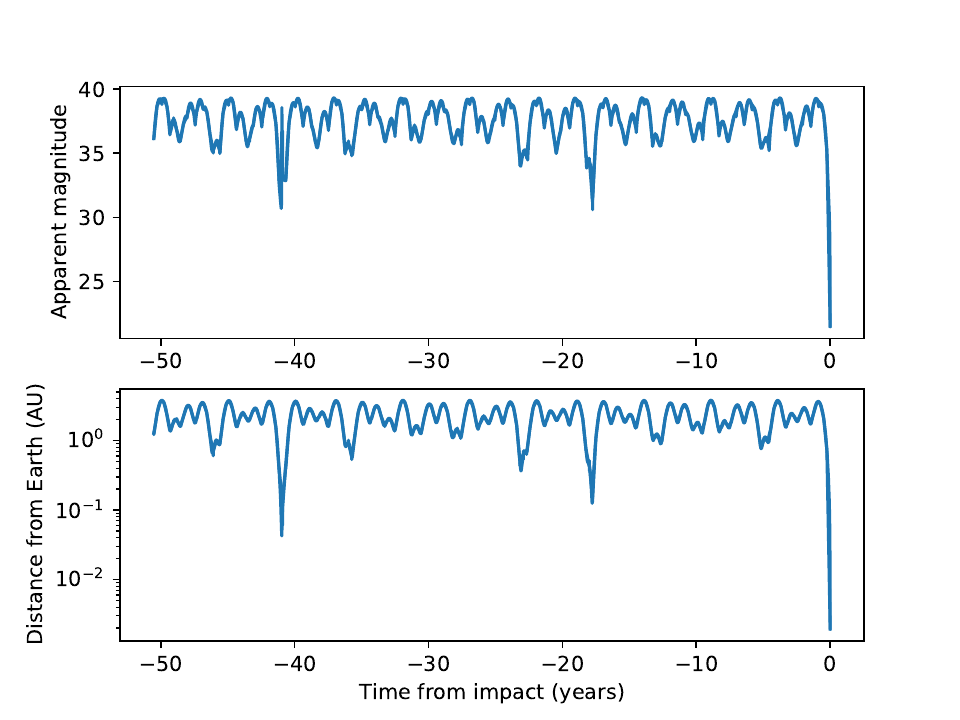}
\caption{The apparent magnitude (upper panel) and distance from the Earth (lower panel) for 2022 WJ1 at one day intervals during the 50 years prior to impact. One hundred clones generated from the JPL covariance matrix area presented; their dispersion is less than the thickness of the plotted lines. WJ1 would have remained too faint to be observed during this time interval. }
\label{fig:visibility}
\end{figure}

Could WJ1 have been detected with a longer lead time? Figure~\ref{fig:visibility} shows the apparent magnitude $m_V$ and distance from Earth $\Delta$ of 100 clones generated from the JPL covariance matrix integrated for the past few decades alongside the major planets but not including non-gravitational effects or higher-order gravitational terms. The brightness estimates for the object use JPL's absolute magnitude $H_V=33.57$ and an assumed slope parameter of $G= 0.15$.

The minimum distance from Earth during this time frame occurred approximately 40 years prior to impact at $\Delta = 0.04$~au but the apparent magnitude $m$ was above 30. WJ1 was at its brightest ($m = 30.7$) a little less than 20 years prior to impact where a favorable phase increased its brightness despite a somewhat larger distance.  These apparent magnitudes would have gone undetected under almost all conceivable circumstances. The two currently most productive near-Earth asteroid surveys are Pan-STARRS, which has a limiting magnitude near 22.7 in $R$ \citep{2013PASP..125..357D}, and the Catalina Sky Survey, whose telescopes have limiting magnitudes from 19.5 to 21.5.\footnote{\url{https://catalina.lpl.arizona.edu/telescopes} retrieved 23 Jan 2024} An additional sensitivity loss due to likely trailing would further complicate pre-covery of an object like WJ1. We conclude that it was all but impossible for WJ1 to have been detected during any close approach in the past 50 years even with survey systems similar to those currently operating.

\subsection{Photometric Approach, Colors, and Size Estimation}

The same LDT observations which provided the astrometry, described in the previous subsection, also facilitated characterization of the object's physical properties. We first needed to decide how to appropriately measure the brightness of such a streaked object, as in a typical image the object was streaked over tens of pixels and thus several seeing widths. We used the \textit{PhotometryPipeline} package \citep{2017A&C....18...47M} to calculate a photometric zero point for every image within the PANSTARRS system based on aperture photometry with a four-pixel ($1.44\arcsec$) radius, about twice that of the estimated seeing. We then manually inspected each image where the streak was detected and estimated a `start' and `end' point for each streak. These estimates were used as starting points for a custom streak photometry code in Python (e.g., not a part of PhotometryPipeline) which first fit for the location of peak brightness of the streak in each column (or group of columns) on the detector and then used these new central values as a function of detector location to determine the location and orientation of the streak. The summed brightness of each of the pixels within the same aperture radius that was used for the background stars ($1.44\arcsec$ from the peak brightness location) was fixed, but the number of pixels \textit{along} the streak which would be used in the sum was used as a free parameter. If one picked a number sufficiently large that the whole streak was summed up, then SNR could be maximized, though at the expense of sensitivity to temporal variations in brightness on timescales less than the exposure time. (Smaller asteroids can have very fast rotation periods \citep[e.g.][]{2024arXiv240404142D}, so this is a loss of potentially interesting information.) That said, choosing to measure the bulk brightness of a streak over a length larger than several stellar PSFs (more than $\sim3-4\arcsec$) did not improve SNR over simply taking the average of several smaller streaks with commensurately better estimations of the local background. Other codes like that of \citet{2012PASP..124.1197V} can work similarly to localize and extract fluxes for streaked Solar System objects, but we developed a new code so as to be able to explore the time-domain brightness variations along the streaks in more detail easier.

At the other extreme, if one chose a smaller length along the streak to extract -- such as the seeing width -- one would have to account for the acceleration of the object across the detector to find the effective exposure time within each segment of the streak. Whatever aperture was used, the brightness values were then corrected to heliocentric and geocentric distances of $1$ AU. Adjustments in along-streak extraction length did not change the median brightness of all of the individual extracted segments.

To assess the accuracy of our extracted brightness values, we converted our distance-corrected Sloan $r$ magnitudes into absolute magnitudes $H_r$ using the same phase curve as JPL, namely an HG curve \citep{1989aste.conf..524B} with G=0.15. Their $H_V = 33.58\pm0.36$ agrees with our average $H_r = 33.93\pm0.03$ to better than $1-\sigma$ after correcting for the Solar $V-r_{PANSTARRS}$ color ($0.17$ mags). Even discounting errors, the two values agree in flux at about the $\sim18\%$ level or better; far smaller than the uncertainty in albedo for likely compositions (see below). We view this as a good indication that our photometric approach was accurate. However, we note that the true phase curve of an object so small is likely different from that of the IAU standard HG system, which was derived from observations of large asteroids likely covered in significant regolith \citep{1989aste.conf..524B}. Asteroids that are too small or spinning too fast (or both) to maintain significant regolith should have phase curves that are significantly different from this HG model. However, we chose to adopt this model rather than choosing another to apply as there is no clear benefit to other models for this event.

\begin{figure}[ht!]
\plotone{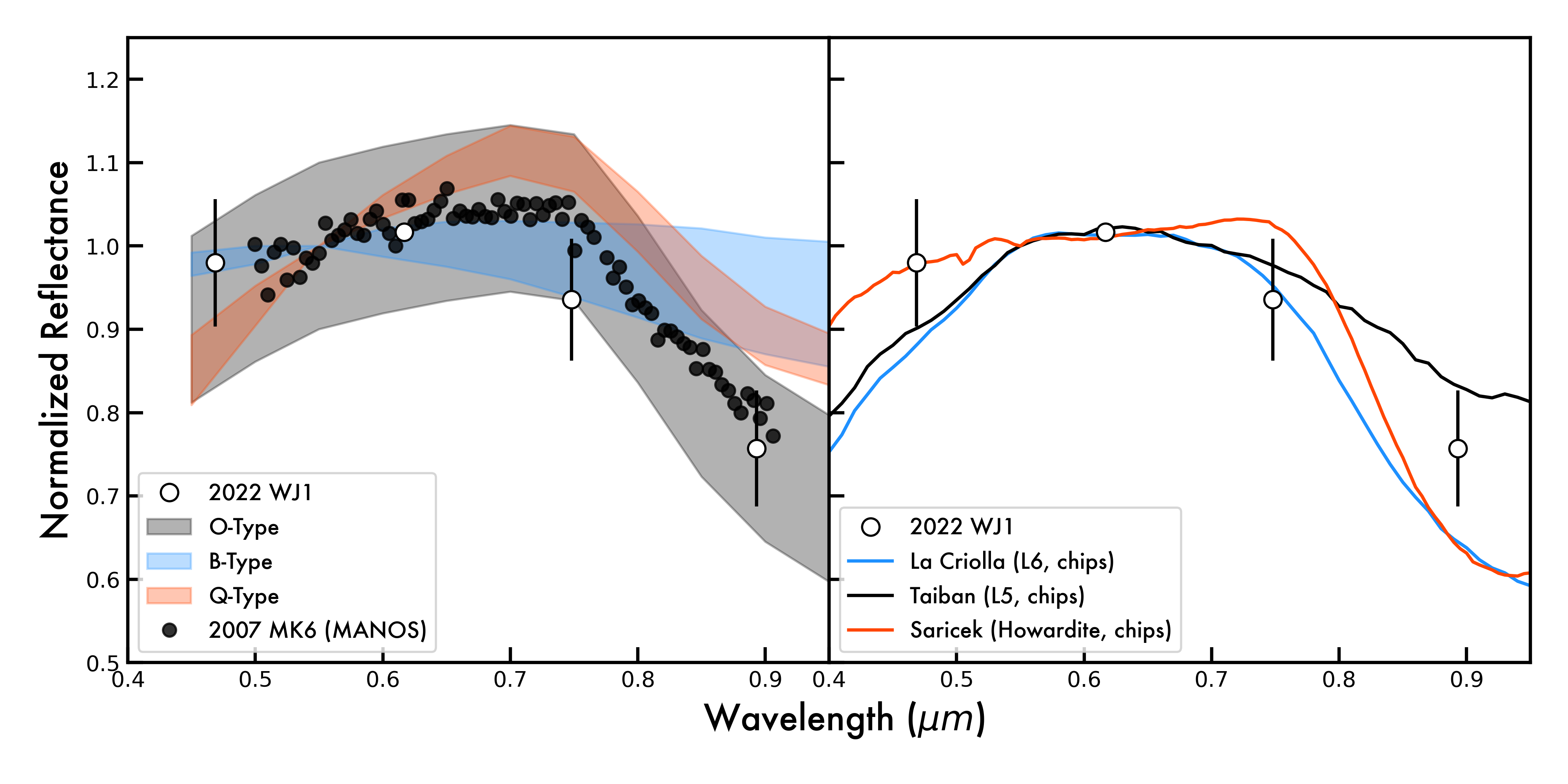}
\caption{A comparison between the photometrically-derived reflectance of 2022 WJ1 (white circles) with reflectance spectra of different asteroid types that share some spectral properties in common (left panel) and with reflectance spectra of meteorites that are potential analogues for those asteroid types (right panel). In the left panel, the O-, B-, and Q-type asteroid classes are plotted as filled areas in grey, blue, and red colors. The vertical spread of these areas indicates the variation of objects within these classes.
The S-Complex asteroid types (particularly the O- and Q-types) are significantly better matches than the B- or other types, indicating that a rocky composition -- and thus moderate albedo and density -- is most likely for WJ1. Given that asteroid taxonomies are based on the reflectivities of larger asteroids, some differences might be expected for a meter-class object like WJ1. The visible spectrum of a $\sim$kilometer-scale O-type asteroid, 2007 MK6 \citep{2019AJ....158..196D}, is also plotted for comparison as black filled circles. While the reflectance of WJ1 does fall within the range of spectral behaviors seen on the O-type asteroids, these size-dependent spectral trends prevent a completely unambiguous association with one rocky type of asteroid over another. In the right panel, the three meteorites plotted, two ordinary chondrites and one Howardite, are spectra of chips of those meteorites \citep{2020pds..data...98M}. They are thus are analogous to a surface with some larger pieces and little reoligth as we might expect for an object as small as WJ1. The composition and reflectivity of 2022 WJ1 are discussed at more length in the text.
}
\label{fig:colors}
\end{figure}

\begin{figure}[ht!]
\plotone{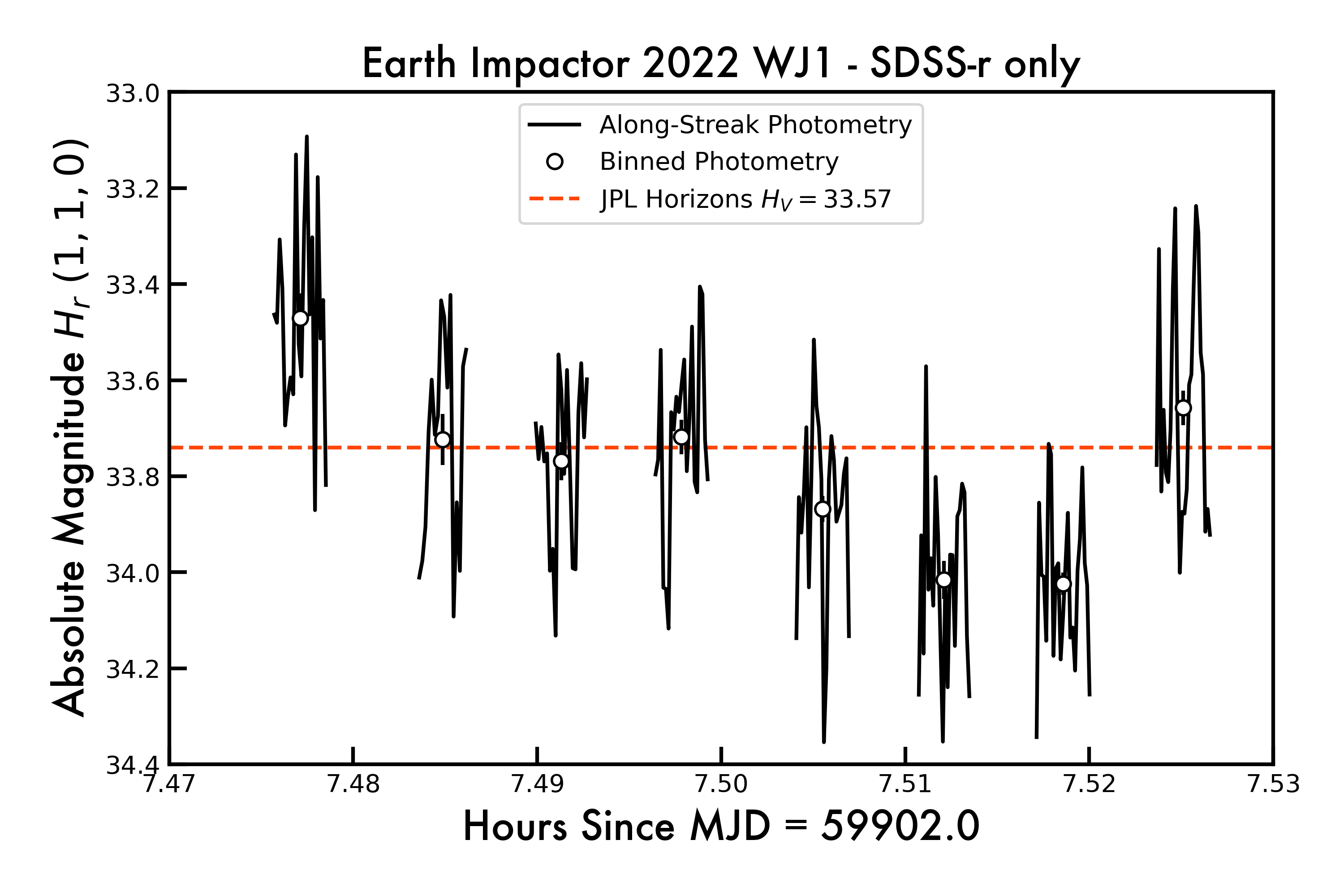}
\caption{The brightness of 2022 WJ1 in the Sloan $r$ filter as measured at the first pointing in which we observed the target. (The full multi-color dataset is presented in the next Figure.) The along-streak photometry is shown as a solid line in black and the median magnitude extracted from that frame as a solid unfilled dot. The absolute magnitude available on JPL Horizons for this object, consisting of dozens of measurements from many observers, is shown as a horizontal red line. The short-term variations along each of the streaks are a combination of sub-second atmospheric variations and the actual changing brightness of the target, while the median magnitudes for each frame vary much more smoothly. That said, and discussed at greater length in the text, no rotation period was found that could fit the brightness variations seen in our full dataset and thus we assume WJ1 to either be tumbling (i.e. non-principal axis rotation) or to have a rotation period too short or too long to resolve with our dataset.
}
\label{fig:r_only}
\end{figure}

\begin{figure}[ht!]
\plotone{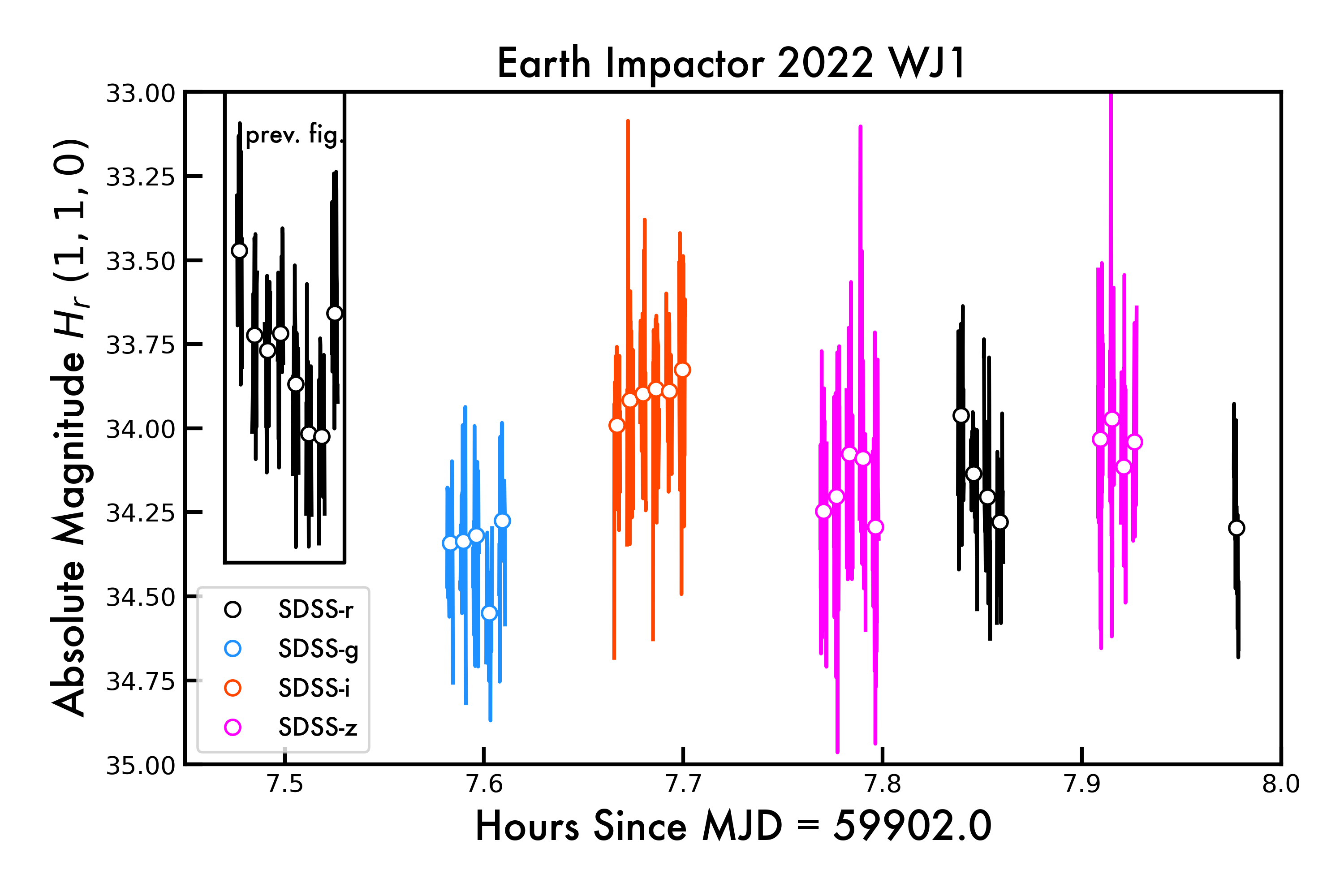}
\caption{The brightness of 2022 WJ1 in the Sloan $g$, $r$, $i$, and $z$ filters. The along-streak photometry is shown as a solid line in each filter's color and the median magnitude extracted from that frame as a solid unfilled dot. The decreasing number of streaks per cluster as a function of time is related to the object's increasing speed on-sky and thus the decreasing time it took for WJ1 to cross the detector. The extent of Figure \ref{fig:r_only} is shown as a box in the top left.
The last observation, a single streak in Sloan $r$, immediately precedes the object passing into the shadow of the Earth and may be artificially dimmer as a result. The effect of the object's placement with respect to the penumbra is discussed in the text.
}
\label{fig:lc}
\end{figure}

To estimate the colors of WJ1 we use  a comparison of the average brightness of the object through each of the Sloan $g$, $r$, $i$, and $z$ filters. The colors of the object were measured to be $g-r = 0.43\pm0.05$, $r-i=0.03\pm0.04$, and $r-z = -0.19\pm0.05$ in the PANSTARRS magnitude system \citep{2012ApJ...750...99T}. We convert these colors to reflectance through comparison with the colors of the Sun \citep{2018ApJS..236...47W}. The results are shown in Figure \ref{fig:colors}. The along-streak brightness measurements for our first pointing in Sloan $r$ are shown in Figure \ref{fig:r_only} to display the general quality of our observations, and all the lightcurves of all four filters are shown in Figure \ref{fig:lc}.

The $g-r$ color of WJ1 is neutral or slightly red, but the $r-i$ and particularly the $r-z$ colors are clearly blue. Also shown in Figure \ref{fig:colors} is a comparison of WJ1's reflectivity against the three best-fit taxonomic types in the Bus-DeMeo system \citep{2009Icar..202..160D} and three spectra of chips of rocky meteorites from RELAB \citep{2020pds..data...98M}. The O-type is clearly the best match (RMS $=0.055$), followed by the Q-types (RMS $=0.11$) and the B-types (RMS $=0.11$). If only $g-r-i$ colors were collected, the three types would be indistinguishable, but the $i-z$ color is blue enough that a $1-\mu{m}$ band is required in the matched taxonomic type to match the data. The phase angle was stable at $\sim37.6^{\circ}$ throughout the imaging sequence, so no intra-color phase correction was applied and no phase reddening (see, e.g., \citealt{2012Icar..220...36S}) is to be expected. We note explicitly, and discuss at length later in this paper, that the Bus-DeMeo system and other asteroid taxonomies were developed from observations of much larger asteroids. We thus acknowledge that even if WJ1 reflects light like an O-type asteroid, that does not necessarily mean that it has identical surface properties relative to a larger object. We come back to the question of WJ1's probable lack of regolith in the Conclusions section. The meteorite chip spectra were selected as representative (but not best fit) examples of reflectance spectra of rocky meteorites (e.g., Ordinary Chondrites, HEDs) which one would expect to be sourced by S-complex asteroids (e.g., those with significant 1-$\mu{m}$ absorption features) and broadly matched the reflective properties of WJ1. As with the asteroid taxonomies, WJ1 is similar to the comparison spectra but no match is perfect.

In Figure \ref{fig:lc}, we show the `lightcurve' of WJ1 through each of our four filters. Each block of observations is an individual pointing of the telescope, such that each of the solid lines are individual along-streak measurements and the unfilled white circles are the median brightness of those streaks. As can be seen, there are both short term (minute-or-less) and long-term (tens-of-minutes) variations in the object's brightness, but no unambiguous repeating signatures are seen. While changing the along-streak extraction length can alter the brightness variations seen in a single streak, they did not change the median brightness for each frame or the overall average brightness across the whole dataset, as expected. While the colors reported in the previous paragraph are essentially bulk-averages (e.g., the average of all $g$ frames minus the average of all $r$ frames), we did attempt multiple ways to search for a lightcurve (Fourier analyses, Lomb-Scargle techniques, etc.) and use that to correct the colors but found little success. No believable periodic signals were identified through these standard techniques and manually folding the data to various periods could produce interesting results in a single filter but poor results in others (a $\sim35$ s period appeared to phase the $r$ data seen in Figure \ref{fig:r_only} well, in particular). If one interprets the fact that the earlier Sloan $r$ photometry is slightly brighter than the latter photometry in the same filter, and thus that WJ1's rotation state is appreciably longer than $\sim30$ minutes, the short-term variations seen at every pointing become challenging to explain. This does not rule out a short periodicity that we were not able to isolate with the cadence of observations we obtained, but the broad agreement between telescopic and meteor camera data about the nature of the object suggests that any short period patterns in the lightcurve did not significantly limit our ability to determine this object's colors. We thus infer, based both on WJ1's small size and our challenges in identifying a clear periodic signal in its brightness, that the object is possibly tumbling with a short period and is not extremely elongated. We comment on the interpretability of our lightcurve in light of future efforts to characterize Earth Impactors at the end of Section \ref{sec:disc}.

The last frame in which WJ1 was detected (and the last time for which an astrometric position was reported to the MPC) is the final Sloan $r$ observation in Figure \ref{fig:lc}, after which the object was fully lost in the shadow of the Earth. Prior to this, it would be expected that light from the Sun would have been increasingly but non-linearly blocked from the Earth as the object passed through the penumbra towards the umbra but at a rate and with timing that is highly dependent on its inbound velocity and trajectory. This may explain the fact that our last observations in Sloan $r$ are dimmer than the earlier ones. However, considering that our absolute magnitude is consistent (but slightly fainter) than JPL's absolute magnitude which includes observations taken hours earlier, this effect cannot be very large. A simple model suggests that the dimming happens relatively rapidly -- the penumbra is not extremely 'wide' such that a gradual dimming would be expected -- and thus that the object only began to dim appreciably around when we lost it. We also note that including or not including our final Sloan $r$ data changes our average magnitude in that filter by less than our reported error. The two $z$ filter pointings also show the opposite trend, which indicates that lightcurve variations could be dominating these brightness changes. If the variability in Sloan $r$ is assumed to solely be due to lightcurve effects (e.g., we treat the variation in brightness in that filter as the whole lightcurve amplitude), we estimate an $a/b$ ratio of $\sim1.5$ or less after accounting for phase angle \citep{2019AJ....158..220L}. As mentioned elsewhere, amplitude variations on smaller timescales than we were able to investigate might influence this estimation. However, the intra-streak variations seen in Figure \ref{fig:lc} as the dark solid lines -- which we think are from a combination of very short-term variations in atmospheric conditions at the observing site (e.g. scintillation noise) and the actual varying brightness of the asteroid -- show approximately the same amplitude, so the true $a/b$ ratio isn't likely to be much larger than what we've estimated through the variation of the average streak brightnesses.

Taking our bulk-averaged colors, modulo these several factors that might have small effect, we conclude that our visible color observations are most consistent with WJ1 having a rocky composition (it has a 1-$\mu{m}$ absorption feature), and thus a moderate-to-high albedo and presumably bulk density. The range of asteroid visual albedos in the S-complex (see, e.g., \citealt{2011ApJ...741...90M}) range from $\sim15-35\%$. From this we suggest that WJ1 had a diameter between $40$ and $60$ centimeters. The stony asteroids are most commonly, though not exclusively, linked to ordinary chondrites, the most common kind of meteorite falls, whose densities fall in the range ($2.5-3.5$ g/cm$^3$) \citep{Flynn2017}. Laboratory studies of the reflectivity of these meteorites suggest that larger particle sizes generally result in lower albedos compared to fine grain sizes \citep{2023PSJ.....4...52B}. As such, a small asteroid with limited or no regolith would be expected to have an albedo on the lower end of the measured range and thus a size slightly above half a meter across. Other rocky meteorites, like the HEDs or the Enstatite chondrites, are also possible candidates to explain the reflective behavior of WJ1. That said, the size-dependent spectral trends expected for the smallest asteroids and discussed throughout this paper make an identification of one type of meteorite over another challenging. While Ordinary Chondrites are discussed more than the other types of meteorites in these analyses, it is only because they are more common and not that they are explicitly the best or only match to the telescopic data.

We note that if the object were instead in the C-complex, a larger size as well and a lower density would be expected than those estimated here. The details of how WJ1 broke up are sensitive to its size and density, and thus we revisit the size and density, and thus composition, in the following subsections related to the atmospheric ablation behaviour of the asteroid.

\subsection{Atmospheric Trajectory}

\begin{figure}[]
\plotone{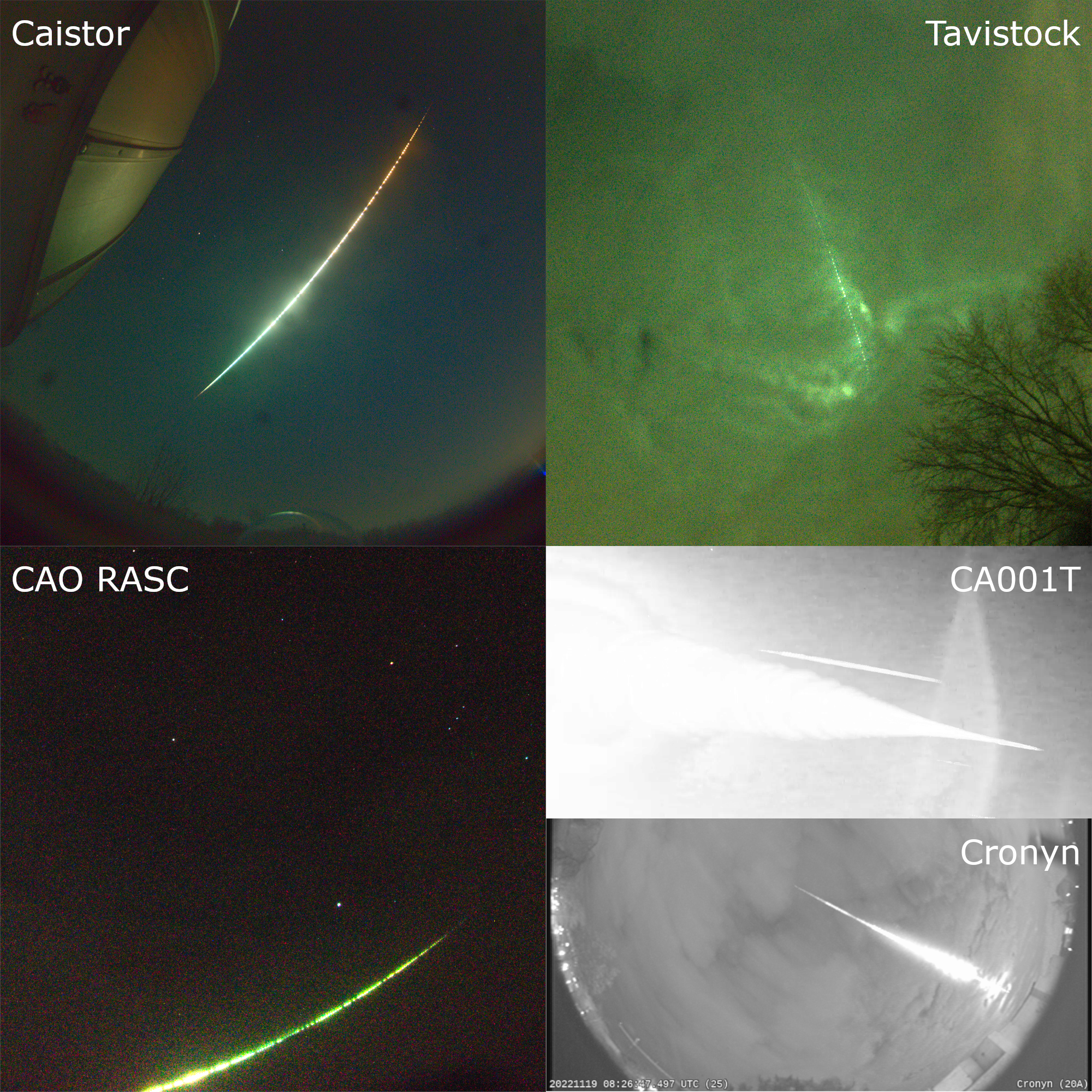}
\caption{Mosaic showing observations of the fireball used in the trajectory solution. The parallel streak in the CA001T image above the bright fireball is due to an internal reflection in the optics. Note the constellation of Orion is visible in the CAO RASC image for scale. The moving direction of the fireball for each camera is towards: Caistor - upper right, Tavistock - bottom right, CAO RASC - bottom left, CA001T - left, Cronyn - bottom right.}
\label{fig:fireball}
\end{figure}

The fireball associated with WJ1 was observed optically by 10 instrumental camera systems operated by the Meteor Physics Group at the University of Western Ontario in London, Ontario Canada. At the time of the fireball, there was high, broken cloud across much of the region, such that some cameras had partially or fully obscured views. From these 10 camera systems, only five cameras with the best geometry and data/sky quality were used in the trajectory solution. The selected instruments include one all-sky camera of the Southern Ontario Meteor Network \citep[Cronyn, ][]{weryk2008southern, brown2010development}, one moderate field of view camera of the Global Meteor Network \citep[CA001T, ][]{vida2021global}, and three all-sky photographic cameras of the Global Fireball Observatory \citep[Tavistock, Caistor, CAO RASC - Carr Astronomial Observatory of the Royal Astronomical Society of Canada][]{devillepoix2020global}. A set of composite images showing the observations is shown in Fig. \ref{fig:fireball}. The list of camera coordinates and parameters are given in Table \ref{tab:camera_coords}.

\begin{table}[]
    \centering
    \caption{Camera parameters and geographical coordinates. F is the astrometric plate scale and the two ranges are the minimum and maximum range to the fireball from the position of the camera. The column Dur. (Sec) gives the total time of visibility of the fireball during which measurements were possible from each station.}
    \label{tab:camera_coords}
\begin{tabular}{llrrrrrrr}
\hline
Location & Network & Latitude (\degr) & Longitude (\degr) & Alt. (m) & F (arcmin/px) & Min. range (km) & Max. range (km) & Dur. (sec) \\
\hline
Cronyn     &  SOMN & 43.00552 & -81.27520 & 253 & 16.5 &  76.2 &  83.1 & 11 \\
CA001T     &  GMN  & 43.57659 & -79.75925 & 203 & 2.5 & 146.5 & 194.1 & 10 \\
Caistor    &  GFO  & 43.08199 & -79.61455 & 202 & 2.0 &  27.8 & 135.3 & 10.8 \\
CAO RASC   &  GFO  & 44.49260 & -80.38361 & 454 & 2.0 & 157.9 & 174.9 & 6.25 \\
Tavistock  &  GFO  & 43.26402 & -80.77215 & 338 & 2.0 &  62.5 &  68.1 & 1.85 \\
\hline
\end{tabular}
\end{table}

\begin{figure}[]
\plotone{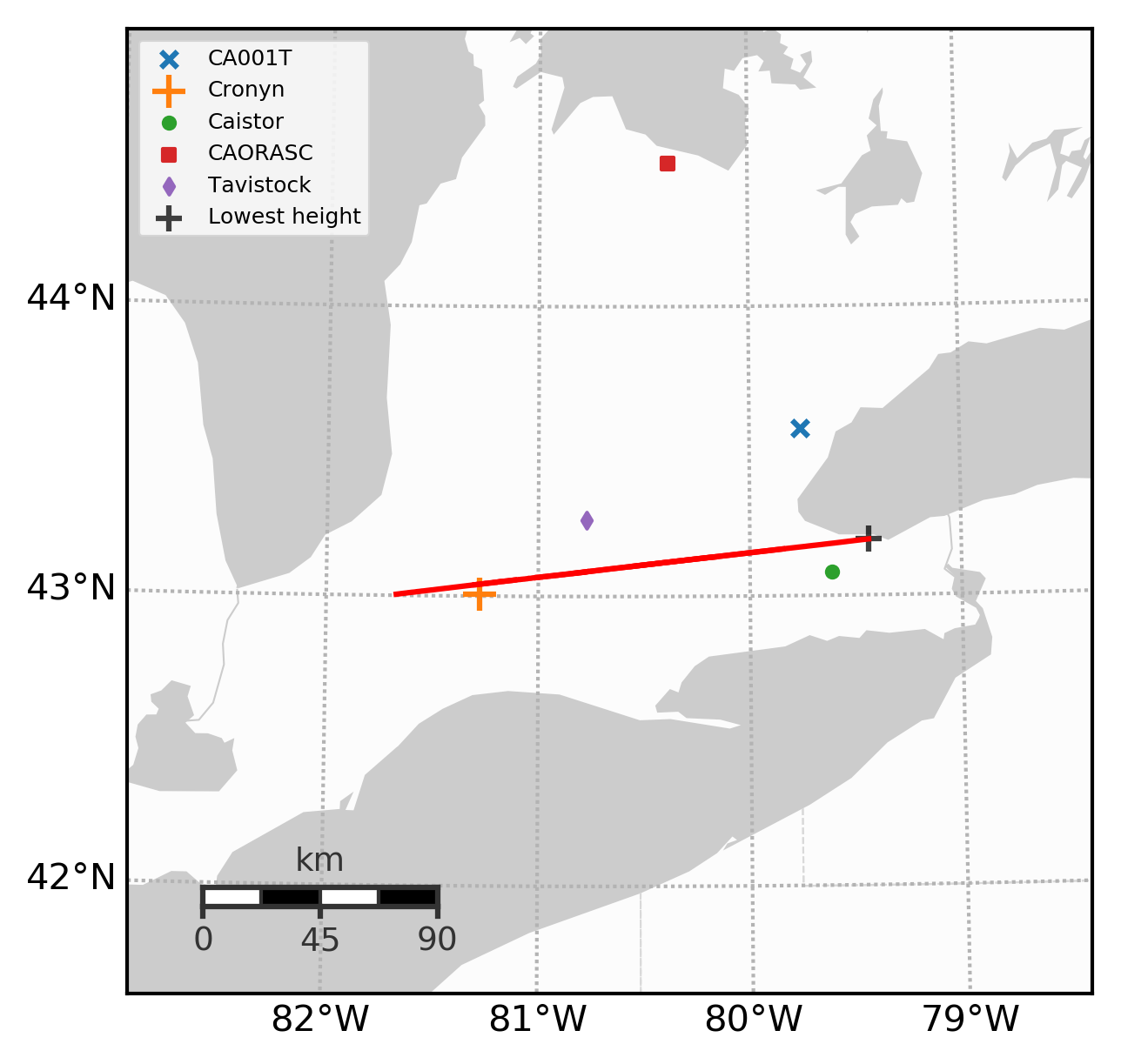}
\caption{The ground track of the fireball (red) together with the locations of cameras used in the trajectory solution. The Great Lakes and other bodies of water are marked in grey.}
\label{fig:map}
\end{figure}

The fireball trajectory was computed using the Monte Carlo time-based trajectory algorithm available in the open source WMPL toolkit\footnote{https://github.com/wmpg/WesternMeteorPyLib} as described in \citet{vida2020}. The ground track of the fireball together with the locations of the cameras are shown in Fig. \ref{fig:map}. The fireball was first observed at a height of 96~km, starting 20~km west of and passing right above the Cronyn camera located on Western University in London, Ontario. It ended at a height of 21.5~km right over the Caistor camera located about 50~km west of Niagara Falls, coming within only 28~km range of that station. The total observed duration is 16.3~s during which the fireball covered a ground path of 204.2~km. The details of the trajectory are given in Table \ref{tab:traj}. We note that despite the SOMN and GFO cameras being time-synchronized, there was a $\pm0.04$~s time offset between the cameras that was compensated for by the trajectory solver. The Cronyn camera was chosen as the absolute reference time.

\begin{table}[]
    \centering
    \caption{Fireball trajectory parameters of 2022 WJ1 determined from ground-based camera measurements. Note that the beginning and end azimuth and elevation are different as the fireball is considered to be a straight line in the Earth-centered inertial coordinate frame, while the azimuth and elevation are given in the ground-fixed coordinate frame.}
    \label{tab:traj}
\begin{tabular}{lrr}
\hline
Parameter & Beginning & End \\
\hline
Time (UTC) & 08:26:40.273 & 08:26:56.474 \\
Latitude (\degr)  & 43.005573     & 43.191669 \\
                  & $\pm$ 25.41~m & $\pm$ 19.80~m \\
Longitude (\degr) & -81.660808    & -79.440835 \\
                  & $\pm$ 58.54~m & $\pm$ 8.24~m \\
Height (km)       & 95.455        & 21.214 \\
                  & $\pm$ 0.044   & $\pm$ 0.013 \\
Velocity (km s$^{-1}$) & 14.003   & $\sim 3.7$ \\
                  & $\pm$ 0.003   & - \\
Azimuth           & 262.697$^{\circ}$     & 264.18734$^{\circ}$ \\
                  & $\pm$ 0.013$^{\circ}$ & - \\
Elevation         &  22.460$^{\circ}$     & 21.03541$^{\circ}$ \\
                  & $\pm$ 0.013$^{\circ}$ & - \\
Geocentric R.A.   &  21.237$^{\circ}$     & - \\
                  & $\pm$ 0.020$^{\circ}$ & - \\
Geocentric Dec.   &  -2.118$^{\circ}$     & - \\
                  & $\pm$ 0.019$^{\circ}$ & - \\
$V_G$  (km s$^{-1}$) & 9.042 & - \\
                  & $\pm$ 0.005 & - \\
\hline
\end{tabular}
\end{table}

\begin{figure}[]
\plotone{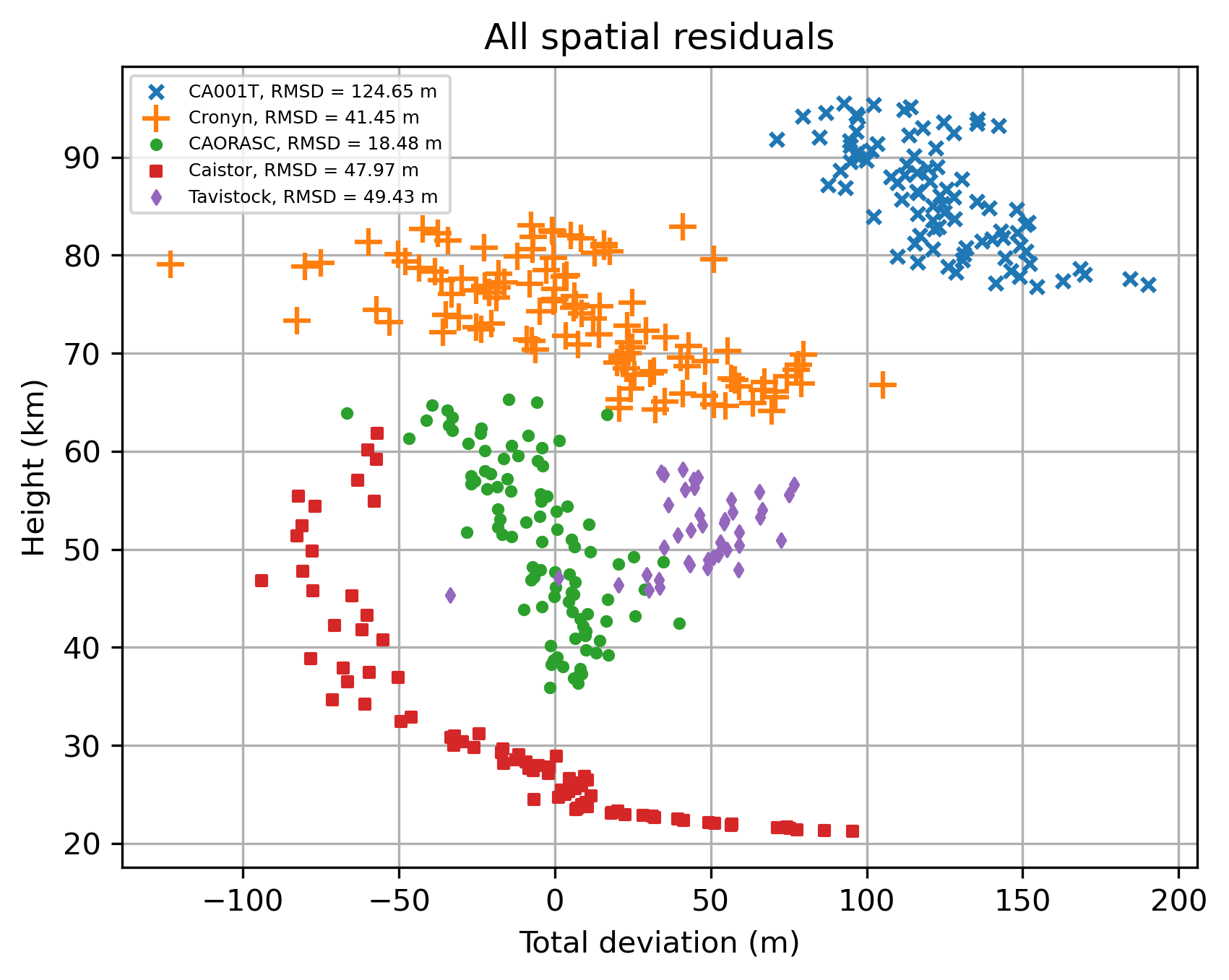}
\caption{The total trajectory fit residuals per camera measurement relative to the best fit trajectory as a function of length from the first observed point.}
\label{fig:traj_residuals_vs_ht}
\end{figure}

Figure \ref{fig:traj_residuals_vs_ht} shows the trajectory fit residuals to the camera measurements. The CA001T, Caistor, and CAORASC cameras recorded enough stars to be calibrated on the image which contained the fireball. The other two captured the fireball through the clouds and an astrometric plate was created using the closest possible recording containing stars, two days after the fireball due to poor weather conditions. The average fit residual errors from all other cameras are well within 100~m. The only exception is the CA001T camera with a notable systematic shift of about 120~m up and left looking down the trajectory. This camera had a low perspective angle to the beginning of the fireball of only 15$^{\circ}$ at a range of $\sim200$~km where it appeared as an almost stationary object, making reliable astrometric picks difficult as the fireball appeared in the video as an undefined blob of increasing size. The Caistor camera also shows a trend which is caused by the saturation of the fireball on the sensor and the non-symmetric point-spread function of the lens. This caused the fireball centroid to shift slightly to one side.

\begin{figure}[]
\plotone{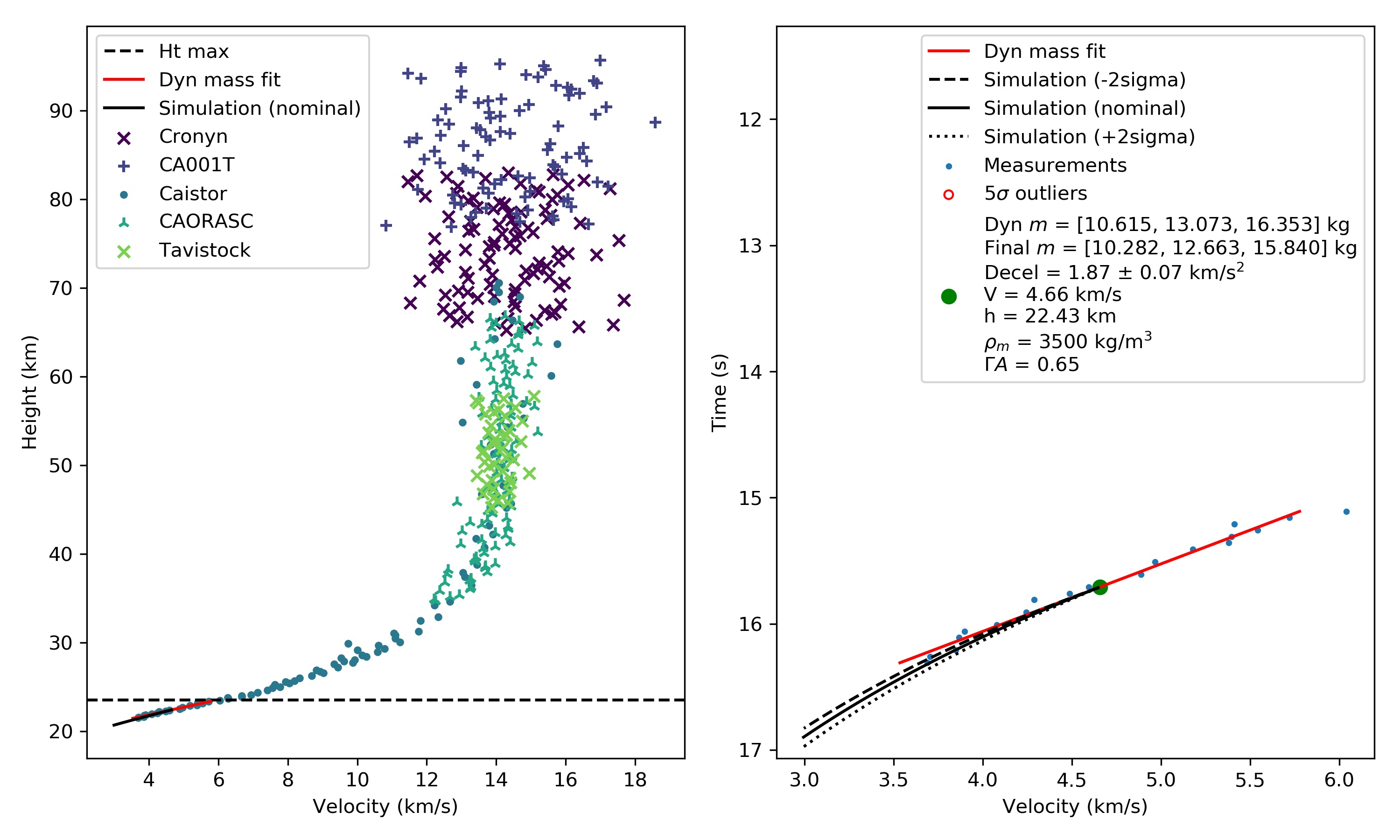}
\caption{Left: Point-to-point velocity measurements for each camera. The horizontal black line shows the cutoff height for the dynamic mass measurement of the meteorite main mass (right). Right: Details of the dynamic mass fit on the velocity measurements - see text for details.}
\label{fig:dyn_mass}
\end{figure}

The Caistor camera also tracked the fireball almost to the ablation limit (based on the recorded end velocity under 4 km/s). This allowed accurate measurement of the velocity and a precise measurement of the dynamic mass of the main meteorite fragment at the terminal point. The dynamic mass of the largest fragment was measured using the method of \citet{mcmullan2023winchcombe}, the details of which are shown in Fig. \ref{fig:dyn_mass}. A classical chondritic bulk density of 3500~kg~m$^{-3}$ has been assumed. A product of the drag factor and the shape coefficient of $\Gamma A = 0.65$ has been chosen for the main mass through a manual empirical approach, with the usual values between 0.5 and 1.0 being tested \citep{borovivcka2020two}. 

The method measures the dynamic mass at a point near the end of the luminous trail, which was chosen as the midpoint of the last 2~km of the observed flight (at the height of 22.4~km and speed of 4.7~km~s$^{-1}$, shown as the large green point in the inset to Fig. \ref{fig:dyn_mass} on the right). The height range has been chosen to match the portion after the last observed fragmentation. The final mass is derived by integrating the single-body ablation equation until 3~km~s$^{-1}$ when ablation is assumed to cease. The 95\% confidence interval for the final mass using the chosen parameters is [10.3, 15.8]~kg, with the nominal value of 12.7~kg. However, we cannot exclude that other similar $\Gamma A$ combinations may work as well. For example, the range of possible masses increases to [8.1, 19.8]~kg for $\Gamma A = 0.65 \pm 0.05$, the range within which the model generally corresponds to the measured velocities.

The fireball has also been captured by numerous security and dash cameras. The first-ever targeted observation of a fireball was made by Robert J. Weryk, resulting in a DSLR photograph of the fireball through clouds. These casual and targeted recordings were not used in the data reduction due to the abundance of instrumental observations. In addition, the fireball has been observed visually by David L. Clark near Brantford, Ontario where initial reports indicated as the entry point, and the co-author Paul Wiegert from a hill at Brescia College in London, Ontario. To our knowledge this was only second time that visual observers had ever been cued in advance to observe a predicted fireball, after two KLM pilots saw 2008 TC3's entry into the atmosphere from the cockpit of their aircraft, and the first time that a measurement or image was able to be planned in advance.

\subsubsection{Ground track comparison}

\begin{figure}[]
\plotone{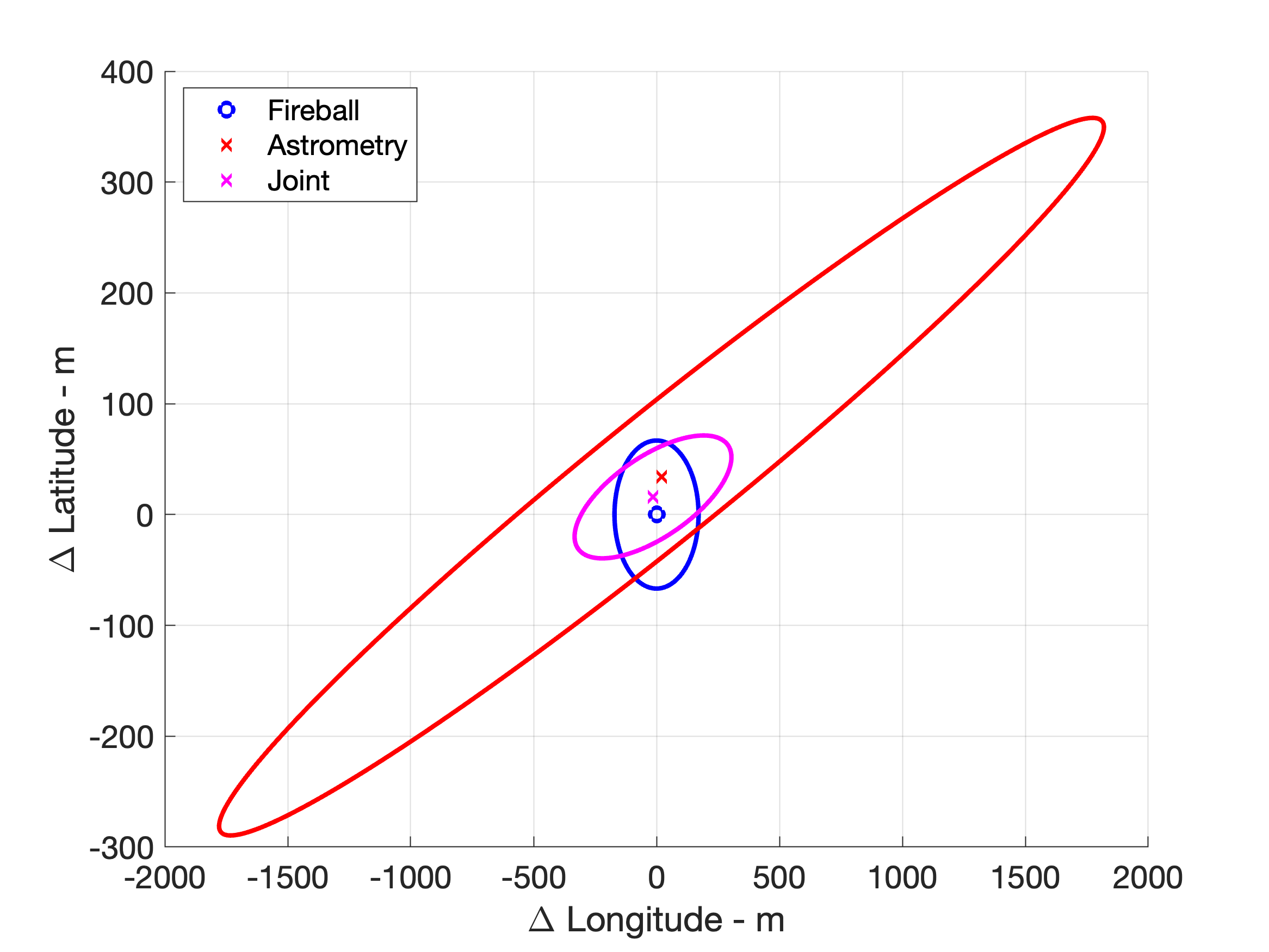}
\caption{Comparison between the ground-based entry point (``Fireball"), telescopic-only solution (``Astrometry") and the combined solution (``Joint") at a reference height of 95.455~km. The footprint ellipses represent 3$\sigma$ errors. The telescopically-estimated ellipses are oriented in the direction of trajectory propagation, with the telescopic-only solution having a major/minor 1$\sigma$ length of 0.609/0.024~km and the combined solution 0.107/0.014~km.}
\label{fig:impact_point}
\end{figure}

The telescopic orbital solution allows calculation of the point of entry of the asteroid into the atmosphere. Figure \ref{fig:impact_point} shows three points: the first observed point on the ground-based trajectory at the height of $95.455\pm0.044$ km, the point of the telescopic trajectory projected to that height, and a combined solution. The combined solution uses the coordinates and timing of the ground-based entry point as a constraint in the telescopic solution. The values for each point are listed in the Table \ref{tab:impact_points}. The impact point calculated through forward propagation of the nominal orbital solution to the reference height differs from the observed in-atmosphere reference point by only 40~m, well within the $\sim600$~m 1$\sigma$ uncertainty ellipse. The combined solution with an uncertainty ellipse of 108~m differs only by 22~m.  Note that the in-atmosphere reference point is high enough that deceleration due to drag remains negligible. This demonstrates that even such a short lead time detection as was the case for WJ1 allows for a well-localized impact location.  Subsequent modelling of the motion at lower heights depends on harder-to-quantify atmospheric interactions including ablation and drag during the fireball phase, and winds during the dark flight phase, which increase the uncertainty in the extent of the meteorite strewnfield.

\begin{table}[]
    \centering
    \caption{Comparison between coordinates of the fireball entry point observed using ground-based cameras, telescopically, and by combining the two. All points are evaluated at the height of $95.455 \pm 0.044$~km. The differences are given relative to the ground-based point.}
    \label{tab:impact_points}
\begin{tabular}{lrrr}
\hline
Parameter         & Ground-based  & Telescopic & Combined \\
\hline
Latitude (\degr)  & 43.005573     & 43.005880  & 43.005716 \\
Longitude (\degr) & -81.660808    & -81.660559 & -81.660984 \\
$\Delta$ (m)      &   0           &  40.3      &  21.7 \\
\hline

\hline
\end{tabular}
\end{table}

\begin{figure}[]
\plotone{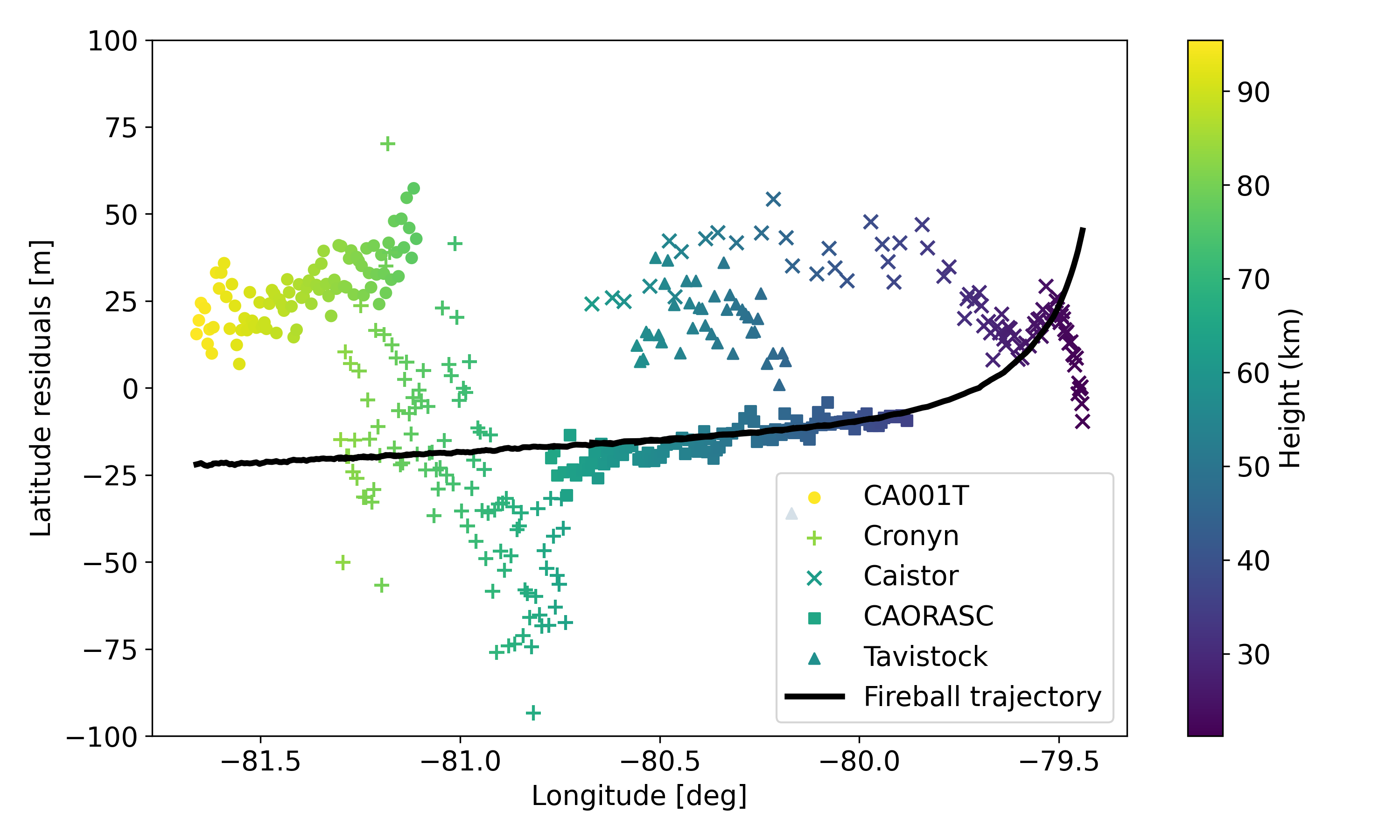}
\caption{Difference in latitude between the ground tracks of the fireball and the asteroid trajectory. As the fireball travelled almost directly due east, only the difference in the latitude (cross-track) is shown. The two data sets are shown: (a) the black curve shows the difference between the fireball trajectory and the asteroid trajectory; (b) the height color-coded points show the difference of camera measurements with the asteroid trajectory.}
\label{fig:ground_track_comparison}
\end{figure}

Figure \ref{fig:ground_track_comparison} shows the comparison of cross-track differences between the telescopically-determined asteroid and fireball trajectory. The fireball was moving from west to east, thus from left to right in the plot. The difference between the two trajectories is well within measurement uncertainties, as indicated by the scatter in the camera astrometric picks. Note that the difference would be significantly higher if the heights were compared directly, as the asteroid trajectory model does not include atmospheric drag at lower heights, which is the cause of the sharp curving in the difference in the fireball's trajectory near its end.

We note that the compensation for the bending of the fireball's trajectory due to gravity was essential in the accurate reconstruction of the trajectory. Without it, the reference point of the fireball's trajectory is $\sim750$ m offset to the east compared to the asteroid trajectory solution. We also roughly investigated whether any lift was present by applying a range of modifying factors in front of the gravity correction parameter (essentially dampening the bending of the trajectory due to gravity). We found that the best fit is achieved with the full correction for bending due to gravity, with even a 10\% dampening showing noticeable offsets. The trajectory measurement accuracy was not sufficient to investigate even smaller damping factors.

\subsection{Fireball Light curve}

Measuring the fireball light curve was the most challenging aspect of characterizing the WJ1 fireball, as all cameras that recorded the fireball either saturated after the initial rise in brightness or captured the meteor through clouds. The fireball was also not registered by the Geostationary Lightning Mapper \citep[GLM; ][]{rumpf2019algorithmic, ozerov2024goes} instrument aboard the Geostationary Operational Environmental Satellite (GOES) even in the level zero data, setting the upper limit on the brightness to about absolute magnitude (normalized to 100~km range) $-14$ \citep{jenniskens2018detection, vojavcek2022oxygen, Wisniewski2024}.

The light curve was derived using a single all-sky SOMN camera at Caistor which was not used in the trajectory solution but which captured the fireball in its entirety. The brightest part of the fireball was saturated. The only unsaturated parts were the first three and the last one second of the total 16 seconds the fireball was observed. The light curve of the saturated portion was corrected by applying an empirical correction derived from laboratory measurements of camera saturation using a calibrated light source following the same general procedure described in \citet{Kikwaya2010}. 

The validity of the derived light curve was confirmed by comparing all unsaturated measurements of the fireball from all cameras. Not including two short flares lasting 1-2 video frames (30 - 60~ms), the brightest portion of the fireball never exceeded magnitude $-14$, consistentwith the non-detection by GLM. Several other cameras caught parts of the light curve without saturating to about magnitude $-9$, and we confirmed that those parts match the measurements from Caistor to within $\pm0.2$ magnitudes. However, parts of the light curve brighter than magnitude -9 might have errors of up to $\pm1$ magnitude.

\subsection{Fragmentation Modelling}

The fireball fragmentation modelling is based on the observed light curve and the fireball dynamics. It has been performed using our implementation \citep{vida2023direct, vida2024first} of the \citet{borovivcka2013kovsice} semi-empirical fragmentation model and has been applied effectively to many meteors before of various compositions \citep{borovivcka2015instrumentally, brown2023golden, mcmullan2024winchcombe}. In this model, meteoroid ablation is assumed to proceed mainly through fragmentation caused by a release of discrete fragments, either via the splitting of the main body or the ejection of mm-sized and smaller dust. Each ejected fragment is numerically integrated using the classical single-body ablation equations. Critically, the model introduces the concept of eroding fragments, i.e. fragments which lose mass both through thermal ablation and the continuous release of dust grains from the fragment surface after which each grain ablates independently.

Table \ref{tab:model_phys} summarizes the best-fit global physical and dynamical parameters of the meteoroid based on the model comparison to the data. In addition, Table \ref{tab:fragmentation} presents a version of the modelled fragmentation behaviour needed to explain the features in the light curve. We emphasize that this is not a unique solution but simply representative. The fundamental issue with the model is its non-linearity and complexity. Defining the uncertainties using this model is a focus of intense study and some recent progress has been made using machine learning algorithms \citep{henych2023semi}. The general fitting procedure we adopted is described in more detail in \citet{vida2024first} where it has been successfully applied.

\begin{figure}[]
\plotone{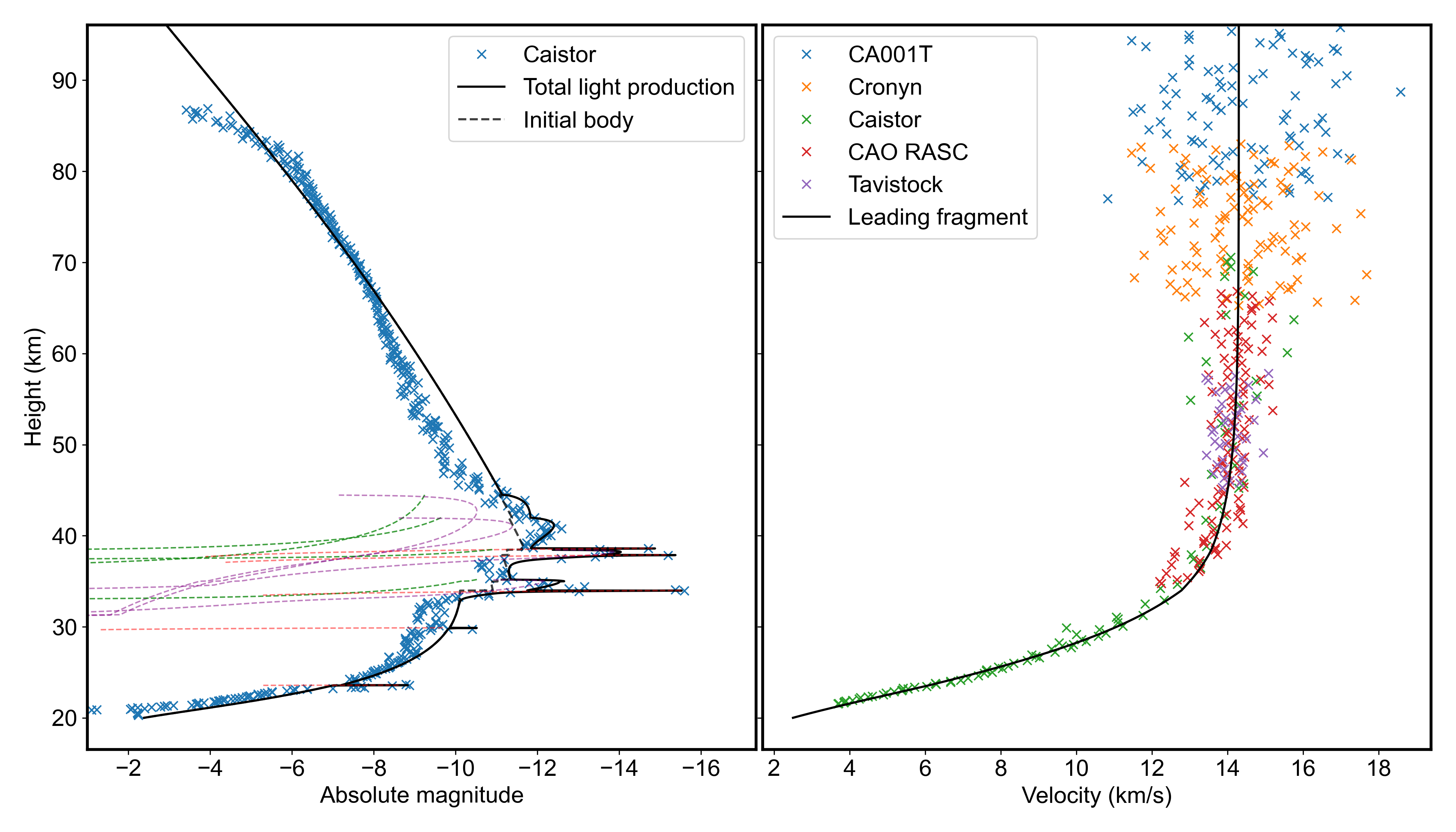}
\caption{Left: Measured fireball lightcurve as a function of height from the Caistor camera (blue crosses) as compared to the total light production estimated from the semi-empirical model (solid black line). The individual lightcurves for eroding fragments (green dashed line), dust (red dashed line) and dust released from eroding fragments (purple dashed line) is also shown. The modelled light production from the ablation of the main mass is given by the dashed black line. Right: The measured point to point velocities for each camera as compared to the model estimate of the velocity for the main (leading) fragment as a function of height.}
\label{fig:simulation_comp}
\end{figure}

Figure \ref{fig:simulation_comp} shows the comparison between the model and the observations. The model optimally fits the dynamics and successfully reproduces the main features of the light curve - the major fragmentations associated with flares. The model deviates from the light curve in the early portion of the path by about 1~mag around magnitude $-9$. This part of the light curve was at the beginning of camera saturation. We were unsuccessful in fitting these parts while having enough mass to survive the fragmentation episodes. In other words, in order to model the initial stage correctly, we would not be able to model the time of peak brightness well. We interpret this model discrepancy as either the fireball having a lower ablation coefficient than assumed in the model due to some form of preheating or having a different luminous efficiency \citep[as previously observed in many fireballs, e.g.][]{popova2019modelling, spurny2020vzvdar}.  Another possibilty is that the saturation correction was not performing well at the edge of the saturation regime. This initial part is not very important as the bulk of the energy deposition occurred in the brightest parts of the fireball.

\begin{table*}
\caption{Model-inferred physical and dynamical properties of the 2022 WJ1 fireball. Note that the grain density of 3500 kg~m$^{-3}$ and nominal ablation coefficient of 0.005 kg~MJ$^{-1}$ was assumed following \cite{borovivcka2020two}.}
\centering
\begin{tabular}{llrl}
\hline
Description & & Value\\
\hline\hline % inserts double horizontal lines
Initial mass (kg)                              & $m_0$    & 220  \\
Initial speed at 180~km (km~s$^{-1}$) & $v_0$    & 14.300 \\
Zenith angle                                   & $Z_c$    & $65.998^{\circ}$ \\
Bulk density (kg~m$^{-3}$)            & $\rho$   & 3400  \\
Grain density (kg~m$^{-3}$)           & $\rho_g$   & 3500  \\
Ablation coefficient (kg~MJ$^{-1}$)   & $\sigma$ & 0.005  \\
Shape-density coefficient                      & $\Gamma A$  & 0.7   \\
(below 35~km)                                  & $\Gamma$ & 0.6  \\
\hline
\end{tabular}
\label{tab:model_phys}
\end{table*}

The fragmentation behavior of the fireball has been explained by two types of fragmentation: eroding fragments (EF) which cause a gradual rise in the light curve and sudden release of dust (D) which cause flares lasting less than one video frame ($<33$~ms). The mechanism of mass loss from the main body was partitioned between EFs, dust release, and direct ablation as roughly 58\%-30\%-12\%, with each subsequent mode of mass loss accounting for about half the previous one. 

\begin{figure}[]
\plotone{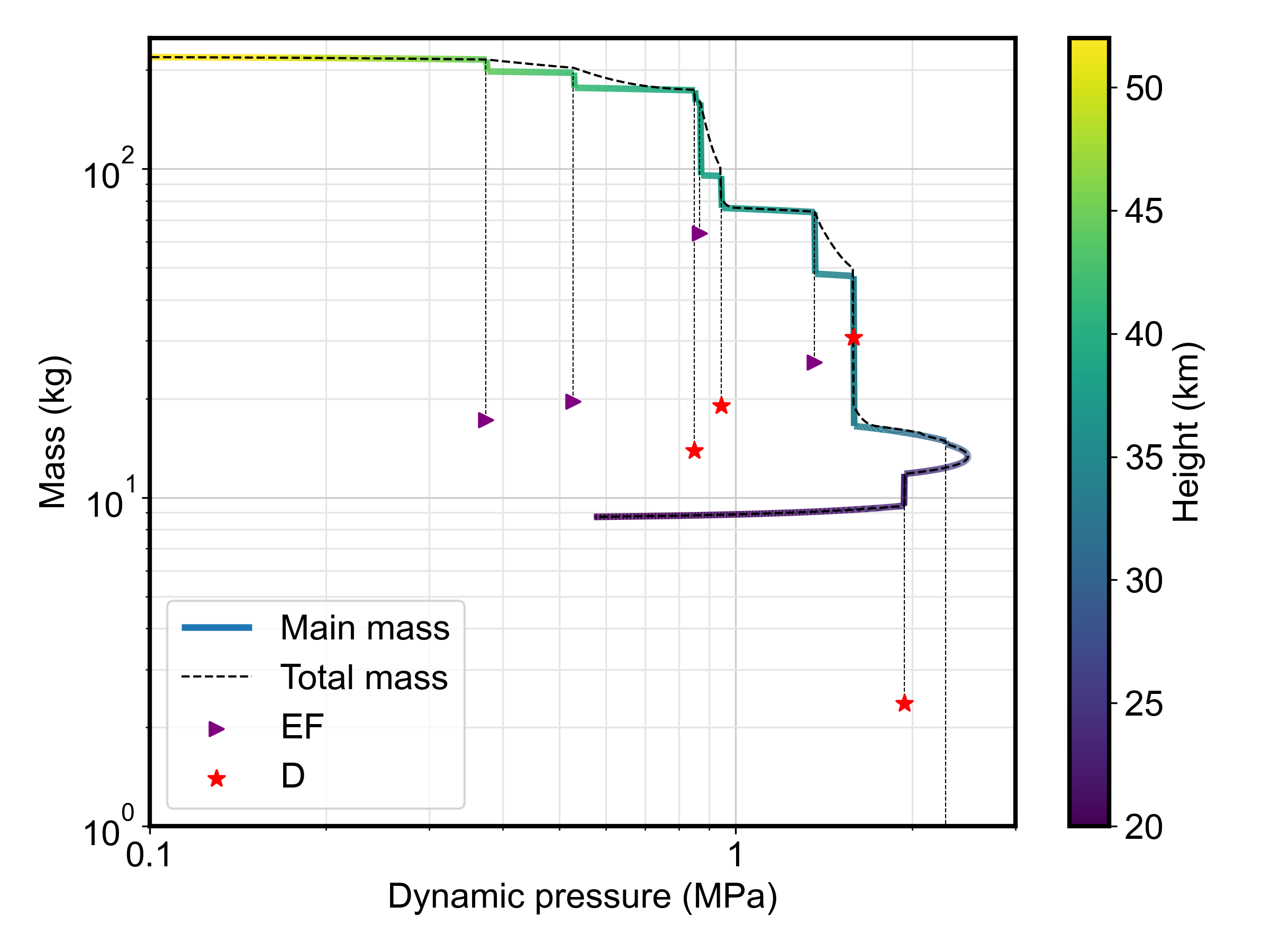}
\caption{The amount of mass remaining in the main fragment as a function of dynamic pressure. This shows the mass loss by fragmentation mode, either leading to a release of an eroding fragment (EF) or dust (D). The fireball height is color-coded. A drag coefficient of $\Gamma = 1.0$ was used to compute the dynamic pressure for consistency with previous work \citep{borovivcka2020two}. The last fragmentation in which 0.3~kg of mass was released into dust is below the lower limit and not shown.}
\label{fig:fragmentation}
\end{figure}

Figure \ref{fig:fragmentation} shows the modelled mass loss as a function of dynamic pressure. Six out of eight fragmentation events caused a similar mass loss of around 20~kg. This is broadly consistent with the results of \citet{borovivcka2020two} for fireballs estimated to have produced ordinary chondrites where the severe second stage fragmentation typically occurred at dynamic pressures of $\gtrsim0.9$~MPa. The final modelled mass of the fragment which survived the atmospheric flight is 8.7~kg, similar to the end fragment dynamic mass estimate. 

Unlike the meteorite-producing fireballs and other fireballs in the \citet{borovivcka2020two} study, WJ1 did not show any evidence for early, first stage fragmentation at low dynamic pressures $\leq0.1$~MPa. This suggests either a lack of regolith and/or matrix material near the surface of the meteoroid. The abundance, or lackthereof, of regolith on the surface of WJ1 is commented on further in Section \ref{sec:disc}.

\begin{table}[]
\caption{Modelled fragmentation behavior. The fragment mass percentage in the table is reference to the mass of the main fragment at the moment of ejection. The mass distribution index for all grains was $s = 2.0$ (see a discussion in \citealt{vida2024first} for how this parameter effects the fit). The values of the dynamic pressure are computed using a drag coefficient of $\Gamma = 1.0$.}
    \centering
    \begin{tabular}{rrrrrrrrrr}
\hline
Time$\mathrm{^{a}}$  & Height & Velocity               &  Dyn pres & Main $m$ & Fragment  & $m$ & $m$     & Erosion coeff            & Grain $m$  \\
(s)   & (km)   & ($\mathrm{km~s^{-1}}$) &  (MPa)    & (kg)       &       & (\%) & (kg)   & ($\mathrm{kg~MJ^{-1}}$) & range (kg) \\
\hline\hline % inserts double horizontal lines
 9.78 & 44.50 & 13.98 & 0.411 & 215.10 & EF &  8.0  & 17.208 & 0.20 & $10^{-4} - 10^{-3}$ \\
10.28 & 42.00 & 13.85 & 0.562 & 196.13 & EF & 10.0  & 19.613 & 0.40 & $10^{-4} - 10^{-3}$ \\
10.96 & 38.65 & 13.59 & 0.849 & 173.36 & D  &  8.0  & 13.869 & -    & $10^{-7} - 10^{-6}$ \\
10.99 & 38.50 & 13.58 & 0.864 & 159.33 & EF & 40.0  & 63.734 & 1.50 & $10^{-5} - 10^{-4}$ \\
11.11 & 37.90 & 13.50 & 0.926 & 95.13 & D  & 20.0  & 19.026 & -    & $10^{-7} - 10^{-6}$ \\
11.68 & 35.20 & 13.05 & 1.264 & 73.83 & EF & 35.0  & 25.841 & 0.40 & $10^{-5} - 10^{-4}$ \\
11.94 & 34.00 & 12.78 & 1.443 & 47.18 & D  & 65.0  & 30.665 & -    & $10^{-7} - 10^{-6}$ \\
12.90 & 29.90 & 11.01 & 1.991 & 14.86 & D  &  2.0  & 0.297 & -    & $10^{-7} - 10^{-6}$ \\
14.99 & 23.60 &  6.12 & 1.795 & 11.80 & D  & 20.0  & 2.361 & -    & $10^{-7} - 10^{-6}$ \\
\hline
\multicolumn{10}{l}{$\mathrm{^{a}}$ Seconds after 2022-11-19 08:26:40.230 UTC.} \\
\multicolumn{10}{l}{$\mathrm{^{b}}$ Final mass of the main fragment at the end of ablation.} \\
\multicolumn{10}{l}{EF = New eroding fragment; D = Dust ejection.} \\
    \end{tabular}
    \label{tab:fragmentation}
\end{table}

As in previous applications of the model, determining the errors on the fit parameters is difficult due to the complexity of the model \citep{henych2023semi}. Here we focus on setting limits on the initial mass, as it bounds the telescopic albedo estimates. (Hence the ablation coefficient and other parameters being assumed to be Ordinary Chondrite-like due to the similarity of fragmentation pressures between WJ1 and those kinds of meteorites, but we comment on the quality of these assumptions below.)
As most of the mass was lost in fragmentations and sudden dust release, the initial mass was very dependent on accurately modelling the flares. For example, one alternative fit we performed which closely follows the observed light curve but ignores flares, resulting in an initial mass of 150~kg. However, we deem this model inaccurate as it does not match the actual lightcurve which shows several clear flares. 

Instead, we compute the range of possible initial masses by considering the range of luminous efficiencies used in the model noting that the luminous efficiency is expected to depend on mass of the ablating fragment/grain. According to \citet{borovivcka2020two}, the luminous efficiency ranges between 2.2\% and 4.3\% for meteoroids with masses of $10^{-7}$ to 250~kg at 13.5~km~s$^{-1}$, respectively. This translates to a range of a factor of two; thus we conservatively estimate a possible range of initial masses for our 220~kg object to be between 150 and 290~kg. Using the assumed meteoroid bulk density of 3400~kg~m$^{-3}$, this translates to a spherical object of a diameter between 44 and 54~cm, with a nominal diameter of 50~cm. 

This is a very similar size range to what was inferred from the asteroidal colors alone, adding confidence to our choice of modeling approach and parameters. That said, should WJ1 have been a different kind of meteorite that has a significant 1-$\mu{m}$ band (such as an HED), the retrieved size would not be significantly different. The lack of fireballs which dropped these other kinds of meteorites whose lightcurves have been modeled like \citet{borovivcka2020two} did for the Ordinary Chondrites, impedes our ability to directly assess those scenarios.

\subsection{Darkflight modelling and the strewn field}

Initial estimates of the strewn field came from the ground track based on the asteroid orbit and Doppler NEXRAD radar observations of a likely debris plume settling following the fireball detected from Buffalo, NY USA. The radar station was only $\sim50$~km away from the fireball's endpoint. The Doppler radar showed clear returns at the expected time and height of fragments/dust released near the end point and time of the fireball with the distribution of returns elongated along the direction of travel of the fireball. These initial ground fall location estimates were very close to the 2009 Grimsby meteorite fall \citep{brown2011fall}, which was also observed by the Buffalo NEXRAD station.

The dark flight was computed using the fireball trajectory, ejecting smaller (1 - 50~g) meteorites from the observed fragmentation points which occurred below 40~km and predicting the fall location of the main meteorite mass under the influence of winds. The Western Meteor Physics Group's Monte Carlo (MC) darkflight model implementation \citep{shaddad2010recovery, brown2011fall} was used for calculations. A chondritic meteorite bulk density of 3500 kg~m$^{-3}$ was assumed and three different meteorite shapes have been tested: spheres (low drag), bricks, and cones (high drag). An uncertainty of 0.5~km~s$^{-1}$ in speed, 0.002$^{\circ}$ in latitude and longitude ($\sim 200$~m), 200~m in height, and 0.1$^{\circ}$ in azimuth and altitude have been used in the Monte Carlo algorithm for darkflight computations to estimate the expected physical scatter in the strewnfield. Table \ref{tab:df_inputs} lists the parameters used for darkflight modelling from each flare and the fireball endpoint.

\begin{table}[]
    \caption{Ejection locations and parameters used for dark flight modelling of fragments released in flares and of the main mass. Time is relative to 2022-11-19 08:26:40.230 UTC. The mass range is informed by Doppler radar observations. The velocity was taken from the fragmentation model.}
    \centering
    \begin{tabular}{rrrrrrrl}
\hline
Time & Height & Latitude & Longitude & Azimuth & Elevation & Velocity & Mass range \\
 (s) &   (km) &    (deg) &     (deg) &   (deg) &     (deg) & (km~s$^{-1}$)& \\
\hline\hline
 10.923 & 38.65 & 43.15190 & -79.96592 & 263.738 & 21.794    & 13.59    & $\sim1$~g   \\
 11.106 & 37.90 & 43.15372 & -79.94291 & 263.755 & 21.776    & 13.50    & $\sim1$~g   \\
 11.838 & 34.00 & 43.16310 & -79.82191 & 263.842 & 21.683    & 12.78    & 1 - 50~g   \\
 12.935 & 29.90 & 43.17501 & -79.66747 & 263.955 & 21.562    & 11.01    & 1 - 50~g   \\
 15.015 & 23.60 & 43.18771 & -79.50085 & 264.078 & 21.431    &  6.12    & 1 - 50~g   \\
 16.300 & 20.81 & 43.19429 & -79.41703 & 264.125 & 21.380    &  3.00    & 1 - 50~g, 5 - 20~kg  \\
 \hline
    \end{tabular}
    \label{tab:df_inputs}
\end{table}

\begin{figure}[]
\plotone{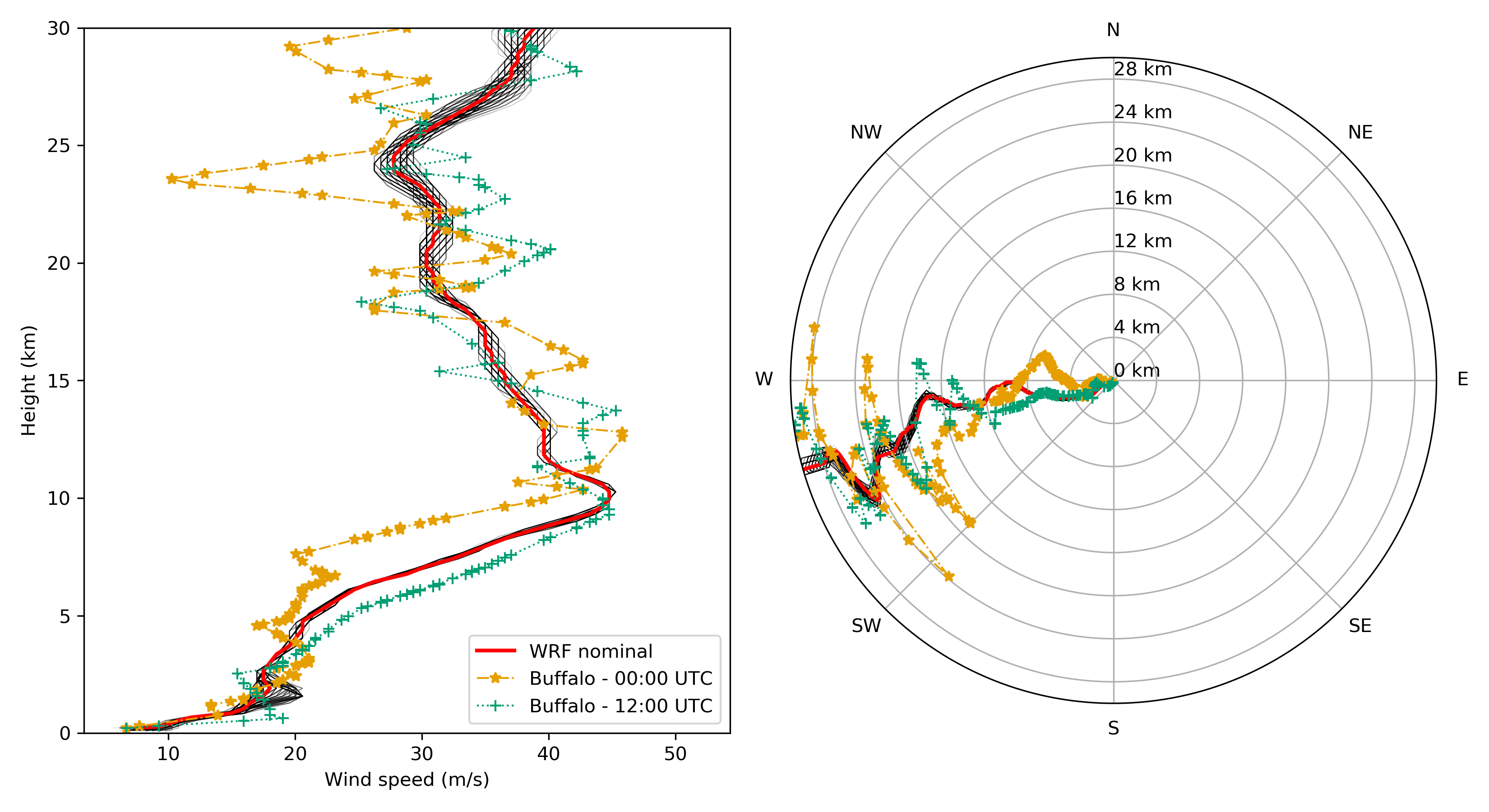}
\caption{Comparison of modelled WRF winds and radiosonde wind profiles. The winds from the WRF model at the terminal point are shown in red and the 400 samples of the model are shown in semi-transparent black lines. The wind measurements from Buffalo on November 19, 2022 at 00 and 12 UTC are also shown for comparison.}
\label{fig:wind_profile}
\end{figure}

The atmospheric conditions and the wind profile were modelled using the Weather Research and Forecasting (WRF) model version 4.0 with dynamic solver ARW (Advanced Research WRF) \citep{skamarock2019WRF4}. The model was sampled up to the height of the highest ejection point and the wind direction, wind speed, air temperature, air pressure, and relative humidity were extracted from the model. Figure \ref{fig:wind_profile} shows a comparison of the modelled winds at the fireball terminal point (red curve) and how they compare to the two closest radiosonde measurements in time from Buffalo, USA (00 and 12 UTC). The Buffalo wind measurements have been retrieved from the University of Wyoming website\footnote{University of Wyoming wind sounding data: \url{http://weather.uwyo.edu/upperair/sounding.html} (accessed Feb 29, 2024)}. The model sampled at the time of the fireball (08:26:40 UTC) most closely matches the measured wind profile at 12:00 UTC, with only minor differences of several meters per second. The winds remained unchanged in direction from 12 hours previously (west-southwest direction), with only the wind speed increasing by about 10~m~s$^{-1}$ in the upper troposphere. In addition, the WRF model has been sampled every 0.5~km in a $10 \times 10$~km window, resulting in a total of 400 samples which are also shown. The plot shows that the winds are very consistent within this temporal and spatial time window and that the same wind profile used for the endpoint can be used for other points of ejection.

For WJ1, estimating an accurate meteorite location was difficult due to the shallow trajectory angle of only $\sim22.8^{\circ}$. This made the strewn field long, amplifying any small errors in drag or winds. Strong winds, which were over 30~m~s$^{-1}$ from the point of ejection down to a height of 7~km and peaked at $\sim40$~m~s$^{-1}$ at the height of 10~km further increased the influence of the unknown shape of meteorites on the final fall locations.

\subsubsection{Fragments - comparison to weather radar observations}

We retrieved the openly accessible Level II data from the nearest US Next Generation Weather Radar (NEXRAD) stations \footnote{NEXRAD on AWS was accessed on Nov 20, 2022 from https://registry.opendata.aws/noaa-nexrad.}.
KBUF (Buffalo, NY), by far the closest station, was only 70km range from the fireball end point.
In the minutes following the fireball numerous returns clearly stood out from the background precipitation and noise, demonstrating once again the sensitivity of NEXRAD systems to meteorite fragments/dust \citep{Fries2010}.
These returns can be broadly grouped into two clusters, 
In the first cluster, some low altitudes returns appeared on sweep 6 (1.7 km altitude) at 08:28:25, 89 seconds after the end of the observed luminous flight.
The subsequent sweeps, made in increasing altitudes, up to sweep 12 (6 km altitude, at 08:29:49) also showed returns.
The second cluster appeared later, when the radar beam intercepted meteorites over a vast area in sweep 17 (14-18 km altitude), at 08:30:52 ($\sim$3 min after the fireball).

\begin{figure}
    \centering
    \plotone{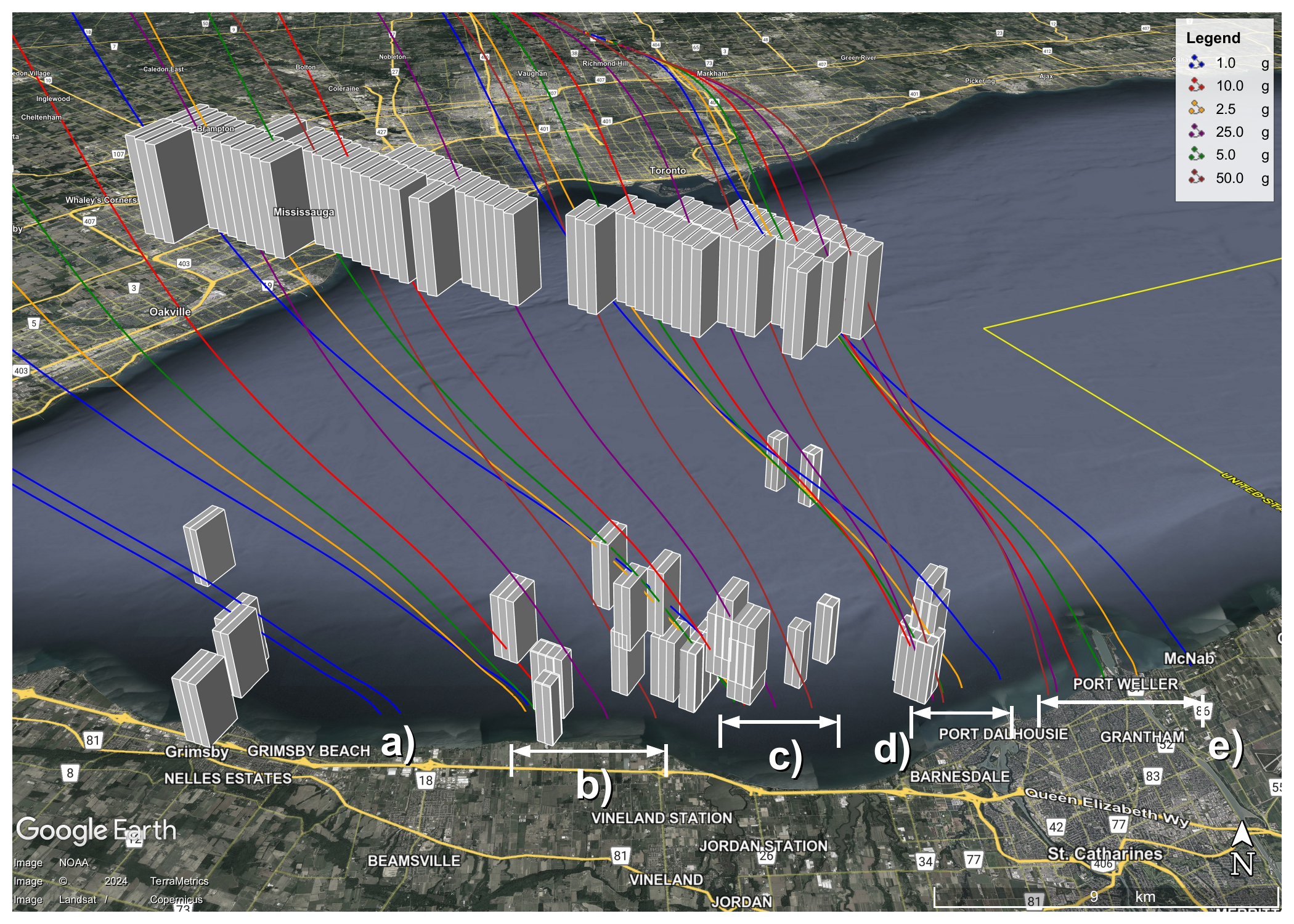}
    \caption{The location and extent of the five doppler radar returns (labelled a through e and which are described in the text) and as well as the dark flight fall curves of small meteorites ejected in fragmentation episodes.}
    \label{fig:df_flares}
\end{figure}

Figure \ref{fig:df_flares} shows the comparison of fall curves of small meteorites ejected in flares to the location of Doppler radar returns. The strewn field for these small meteorites stretches across 27~km on the ground, with almost all rocks falling into Lake Ontario. The assumed uncertainties in the modelling produced a strewn field width of between 1 and 1.5~km, which was not enough for the meteorites to reach land.
As the winds were carrying meteorites towards Lake Ontario, only the best case scenario with the spherical shape (lowest drag) is shown. For all other shapes, meteorites are pushed even further north into the lake. The Doppler radar returns were registered at two discrete slices, with the top having heights of around 15 and the lower one with heights of around 4~km. 

The bottom returns furthest to west (above Grimsby, marked a) in the figure) are best explained in time and location by ejection of $\sim1$~g meteorites from flares at 38.65 and 37.90~km. These two fall curves were very similar and the meteorites passed through the voxels above ground shortly before being blown into the lake. The Doppler returns for these meteorites were only registered at the lowest height slice. The returns marked with b), c) and d) were released from the next three discrete fragmentation episodes (34.0, 29.9, 23.6~km, respectively). The corresponding meteorite fall locations were well separated on the ground. Except for $<10$~g meteorites ejected at 34.0~km which were not registered in the top slice, all meteorites from all three fragmentation episodes were registered at both height slices. 

Based on initial darkflight solutions, a ground search was organized in the hours and days following the fireball event, focusing on Lake Ontario shoreline and immediately adjacent farmland between Grimsby and St. Catharines, Ontario. A public information campaign was undertaken in the region as well, consisting of site visits to homes and businesses, media coverage and distribution of leaflets. Despite the dim prospects for a ground landing based on the NEXRAD and darkflight modelling, it is still possible that some meteorites may have landed southward of the uprange portion of the ellipse. Searching was complicated by regional snowfall and snow clearing of streets during the days after the fireball event. Further searches in the spring of 2023 were also conducted amounting to several hundred person hours, unfortunately with no finds.

Finally, to explain the last $\sim5$~km of the top slice, we ejected 1 - 50~g meteorites from the end height at 20.81~km. Larger $>100$~g masses were outside of the Doppler voxels and probably do not exist. These 1 - 50~g masses are the only non-main mass meteorites that had model fall locations over land. A thorough search of the public spaces and farmland was performed in this area in the weeks after the fall. As most of the strewn field was over the urban area of St. Catharines, Ontario, with single-family homes, leaflets were distributed for residents to be on the lookout for meteorites in their back yard. However, our efforts did not result in any finds.

\subsubsection{Main mass}

\begin{figure}
    \centering
    \plotone{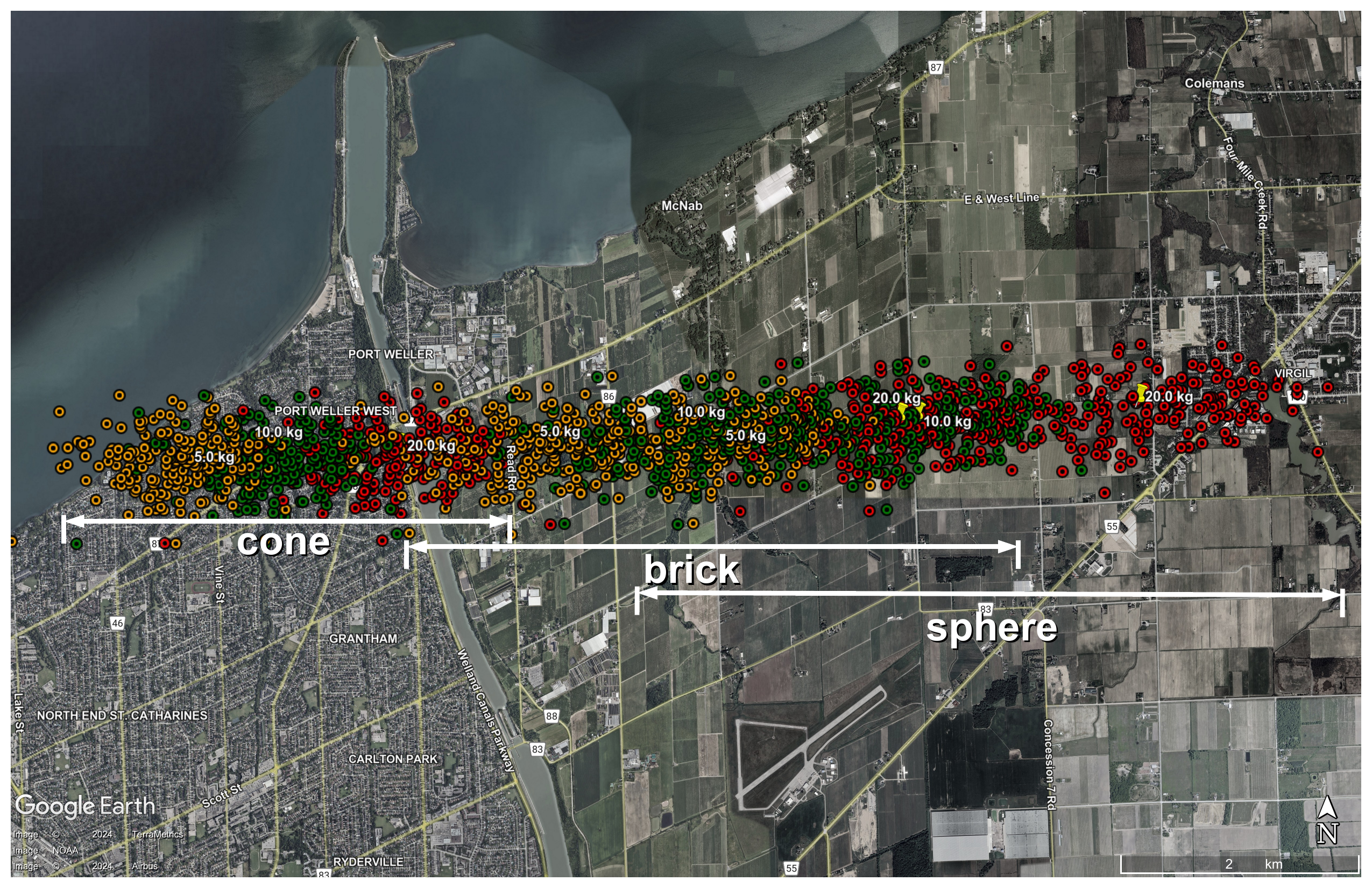}
    \caption{Darkflight model estimate of possible fall locations of the main mass of 2022WJ1 in and around St. Catharines, Ontario. The ground ranges for each hypothesized shape are labelled. Nominal locations for each mass are labelled. Individual points show each Monte Carlo clone in the darkflight simulation (5~kg - yellow, 10~kg - green, 20~kg - red).}
    \label{fig:df_main}
\end{figure}

Figure \ref{fig:df_main} shows the model estimated extent of the strewn field for the main mass. Here we use the fireball end point together with the dynamic mass estimate at the end of luminous flight to inform the expected mass for the main fragment and its fall location. Using  10~kg as the nominal mass and 5 - 20~kg as the extremes we explored several possible meteorite shapes to estimate that the strewn field stretches about 10~km in the east-west direction. Due to the strong winds and the low entry angle, both the shape and the mass have a large influence on the projected main mass fall location. A ground search along the main roads following the center line was performed. Most of the strewn field was private vineyards and visits were made to inform owners of keeping an eye out for a meteorite plunge pit. A drone survey of the area was performed to identify any potential plunge pits, resulting in dozens of candidate locations. As of June 2024, there have been no finds.

\section{Discussion} \label{sec:disc}
The previous sections detailing our investigations into the properties of WJ1 both as an asteroid and as a fireball show good agreement. Perhaps most obviously, this is validation of the utility and intercomparability of the several techniques employed, supporting our conclusion that this is likely the smallest asteroid to have been compositionally characterized in space to date. However, it also facilitates a discussion of where and when these techniques are most useful and what they can tell us about a size of such objects.

\subsection{Orbital Comparisons}
The position and velocity of WJ1 as determined from telescopic and fireball camera astrometry agrees very well (see Figure \ref{fig:impact_point}). Thus the pre-impact orbits one might infer for WJ1 from both techniques also agree very well. The telescopic and fireball orbits have their best-fit ellipses overlap significantly, and their center points are only off by about $\sim40$ meters -- or about $\sim80$ WJ1s across. While the NEO Source Region model of \citet{2018Icar..312..181G} is only formally compatible with asteroids significantly larger (tens of meters instead of half a meter, or $H_V < 25$), that model suggests WJ1 has a $\sim82\%$ chance of having left the Main Belt through the $\nu_6$ resonance. Given that an asteroid as small as WJ1 is almost certainly a fragment of a larger body, if the fragmentation happened relatively recently it is possible that the progenitor object was on a similar orbit and thus the models would be fully applicable. The development of NEO source models which are applicable for smaller objects like WJ1 would add confidence to these kinds of analyses in the future.

This part of the main belt is dominated by stony Ordinary Chondrite like asteroids (see, e.g., \citealt{2014Natur.505..629D}), so a similar composition for WJ1 seems plausible given its orbit alone. This agreement, both between orbit derivation techniques and between what might be inferred for the object's composition from its orbit and from what we actually observed, suggests a broad agreement about the properties of these smaller impactors from multiple techniques.

\subsection{Properties of Ultra-Small Asteroids}
The challenges in detecting meter-scale asteroids with telescopic surveys is that their small size means they can only be found during close approaches to the Earth. Thus the windows to characterize them are short and their sky motion typically large. The bulk of the physical data on objects of this size comes from studying how they break up as meteoroids during atmospheric entry and from laboratory studies of the meteorites that come from and are primarily sourced by them \citep[see, e.g.][]{2016AJ....152..162R}. This means that inferences about significantly larger asteroids typically studied with telescopes are being driven by measurements made of and rocks derived from smaller objects, and thus it is critical to understand how size effects might change the reliability of these conclusions.

Both our telescopic photometry and our modeling of WJ1's lightcurve as a fireball to assess how it broke up in the atmosphere support the conclusion that the object was broadly similar in physical and reflective properties to the stony meteorites and potentially even a specific match to the ordinary chondrite meteorites. That said, the actual reflectivity of the object (see Figure \ref{fig:colors}) is clearly different from larger bodies also linked to ordinary chondrites. This is likely due to differences in their actual surfaces, as smaller asteroids are expected to have a harder time retaining a regolith compared to larger ones -- any process which might remove grains, like rotational instability of the surface or electrostatic lofting \citep{1996Icar..124..181L}, would work more efficiently on bodies with weaker gravity. The lack of a first stage fragmentation of WJ1 during its fireball stage supports the notion that there was little regolith or matrix material in the near surface of the pre-atmospheric meteoroid. 

As mentioned in the telescopic photometry subsection, the typical grain size of an asteroid surface impacts both the overall spectral slope and the relative strength of the absorption features \citep{2023PSJ.....4...52B}. Ordinary Chondrites appear to commonly become more spectrally neutral and to have shallower absorption features as their average grain size increases, though the intensity of these two effects varies with meteorite type \citep{2023PSJ.....4...52B} and is clearly a non-linear effect with grain size. In addition, fresher surfaces (see, e.g., \citealt{2010Icar..209..564G, 2015aste.book..597B}) should have deeper absorption features than more weathered ones and less reddening. Comparing our colors with large-grained samples of ordinary chondrites \citep{2023PSJ.....4...52B} shows good agreement given the differences in spectral resolution. While our colors are not sufficiently precise to be able to quantitatively infer a grain size, we can say that a comparison to laboratory data supports the scenario that WJ1 reflected light as if it were monolithic (e.g., a single rock) or covered in large particles or chips as opposed to being covered in small particles. This is most consistent with WJ1 having limited or no regolith as described above and with the lack of a low dynamic pressure, early stage fragmentation during the fireball phase of flight. 

If we had collected data over a narrower wavelength range, an assessment of how to interpret the object's colors would have been more challenging. Given that C-complex objects mostly have neutral visible slopes, our $g-r-i$ colors alone would have been insufficient to discriminate between a S-complex object without much regolith or any kind of C-type body. One might expect then that a larger sample of ultra-small asteroids characterized strictly through visible wavelength data might show an overabundance of C-complex bodies compared to what is actually found in new meteorite falls if these kind of effects are not considered. We thus recommend that future efforts to study the properties of these ultra-small asteroids (or at the very least those with very short observing windows) make every effort to prioritize the broadest wavelength coverage possible. In the case of truly limited time to characterize an object, the $i-z$ color might be the best choice as it allows the observer to discriminate between objects with and without a $1-\mu{m}$ absorption feature driven by the presence of silicates olivine and pyroxene.

The visible wavelength telescopic spectrum used to classify Earth impactor 2008 TC$_3$ as an F-type asteroid (a C-complex subtype with limited or no decrease in reflectance at shorter wavelengths) can thus be understood in more depth. The fit between the telescopic data and the laboratory spectra taken of the meteorites after the fact is quite good, but there are notable differences between the telescopic spectrum and that of a typical F-type asteroid -- the asteroids used to construct the Bus taxonomic system \citep{2002Icar..158..146B} were all significantly larger and thus likely had different kinds of surfaces. The existence of laboratory data in that case and meteor observations in our case helped to break these degeneracies and ensure the reliability of conclusions, but colors or spectra alone might not be enough. Future multi-wavelength observations of ultra-small asteroids and Earth impactors might occasionally retrieve reflective information about an object that actually does little to discriminate between compositions. In other words, mismatches are to be expected until more laboratory studies are completed on the effects of grain size and until more ultra-small asteroids are characterized.

\subsection{Future Observations of Ultra-Small Earth Impactors - Lessons Learned}
Our study was unable to constrain the rotational state of WJ1 prior to impact for multiple reasons. Our coverage of the objects brightness had significant gaps due to changes in filters and the `point-and-wait' strategy of observing the object as it moved through consecutive starfields, so perhaps these gaps made identification of periodic signals more challenging. That said, asteroids as small as WJ1 are significantly more likely to be in a rapid rotational state, so this would require WJ1 to have a rotational period similar to the cadence of gaps in the data. Some asteroids this small are also in a tumbling rotational state, in which case no repeating signals would be expected, but we cannot explicitly confirm or refute a tumbling rotational state at this object.

As noted previously, tracking at the objects rates was not possible due to the telescope explicitly requiring a file formatted to look like the output from JPL Horizons as opposed a file from JPL Scout where new potentially interesting discoveries are listed. While this specific issue has now been addressed at the LDT, time \textit{was} lost as these issues were navigated and fixed in real time. Future observers interested in Earth Impactor research might consider doing a `dry run' of this kind of event at their facilities to shake out these kinds of formatting problems. If we hadn't had this issue and had been able to acquire the object and track at its non-sidereal rates, we would have attempted to acquire visible spectra instead, so the quality of information we would have been able to obtain about the composition of the object would have been significantly improved. Even if that were not possible, we could have obtained a lightcurve with fewer gaps which in turn might have helped determine its rotational state more fully.

WJ1 does not appear to have been very elongated with an estimated $a/b$ ratio of $\sim1.5$ or less, but this is uncertain as we are not sure if we have sampled the full lightcurve or not, or if $100\%$ of the variation seen is due to the lightcurve alone and not due to ambient atmospheric variations on the timescale of a typical streak. If the rotational amplitude had been higher, deriving reliable colors from a functionally unknown lightcurve would have been significantly more challenging. Without a clear understanding of the object's rotational brightness variability, variations in the brightness seen through other filters is significantly harder to interpret. In this sense, the fireball camera data supporting an Ordinary Chondrite like composition provides good evidence that our color-lightcurve correction was reasonable.

This highlights the inherent benefit that spectral observations, even those taken at very low resolution, have over photometric observations -- no lightcurve correction is required. One path forward for Earth impactors which are too faint ($m_V > 20$) or poorly placed for spectroscopic characterization (no available telescopes with ideal instrumentation) would be to employ different observing strategies to `get around' some of the issues with lightcurve corrections. Instruments capable of simultaneous imaging in multiple filters, like Muscat \citep{2020SPIE11447E..5KN} on the 2m Faulkes Telescope North or the upcoming SCORPIO on Gemini South \citep{2022SPIE12184E..68V}, would not require this kind of correction at all. It might be possible to use data from smaller aperture telescopes which focus on obtaining a high-cadence lightcurve through a wide bandpass filter. This would allow larger aperture telescopes to use multiple filters and still be able to correct their data in a reasonable fashion should they not be able to observe multiple wavelengths at once, though this would introduce other telescope-to-telescope calibration concerns. All of the above pre-supposes significant coordination in data gathering either between facilities or within an observing complex, a task which will only be successful if planned before the next event. 

\section{Conclusions} \label{sec:con}

We have compared telescopic measurements of the short-arc impactor 2022WJ1 with ground-based fireball measurements of the same object. Our major findings include:

The colors of the object were measured to be $g-r = 0.43\pm0.05$, $r-i=0.03\pm0.04$, and $r-z = -0.19\pm0.05$ in the PANSTARRS magnitude system \citep{2012ApJ...750...99T}. We derived an absolute magnitude of $H_r = 33.93\pm0.03$, which is consistent at the $1-\sigma$ level with JPL Horizon's absolute magnitude $H_V = 33.58\pm0.36$ after accounting for the $0.17$ magnitude brightness difference of the Sun between those filters. 2022 WJ1 had a telescopic reflectance spectrum most consistent with a silicate-rich asteroid. Together with the fireball fragmentation behaviour and orbital origin likely from the $\nu_6$ escape region, this supports an S-class identification. Considering the relevant range of albedos for S-complex asteroids ($p_V\sim0.15-0.35$, \citealt{2011ApJ...741...90M}), we infer a diameter of $D = 0.4 - 0.6$ m from the telescopic data. No consistent rotation period was visible in the telescopic data, suggesting either tumbling rotation or a short/long period outside the measurable temporal resolution of the time sampling.

Based on modelling of the fireball entry, a best estimate for the pre-atmospheric mass is 220 kg with an estimated range given systematic model uncertainties of [150, 290] kg. For an assumed chondritic bulk density of 3400 kg~m$^{-3}$ this corresponds to an object between $D = 0.44 - 0.54$ m in diameter. Using the absolute magnitude value of $H_V = 33.58\pm0.36$ this translates into a p$_v = 0.27\pm0.05$, also consistent with an S-complex asteroid and in agreement with what was estimated from the telescopic color taxonomic comparisons above.

The telescopic and fireball determined trajectories agree to within their respective uncertainty ellipses, showing offsets of 40m at the reference height of 95.455 km. This is the first detailed telescopic to fireball trajectory comparison and provides direct validation of the meteor trajectory estimation procedure of \citet{vida2020}. Based on the deceleration of the fireball near its terminal point, the surviving major fragment had a dynamic mass of $12.7\substack{+3.1 \\ -2.4}$ kg.

The fireball showed no significant fragmentation below a dynamic pressure of 0.3 MPa, with the majority of fragmentation mass loss occurring for dynamic pressures above 0.8 MPa. This is similar to the second stage fragmentation observed for most chondritic-like fireballs \citep{borovivcka2020two}. The lack of an early, first stage fragmentation suggests little regolith or matrix was present in the upper layers of the asteroid. This might explain why the reflectance spectrum of 2022 WJ1 is somewhat different than larger stony asteroids with regolith on them.
 
Darkflight modelling indicates that almost all fragments landed in Lake Ontario. The main mass, characterized well by the final stage of luminous flight as detected by the sensitive video camera at the nearby Caistor station, should be on land. Despite extensive searches, no meteorites have been recovered as of the summer of 2024, but residents in the area near St. Catharines, Ontario are encouraged to continue looking for what we believe is the main fragment, likely embedded in the ground.

Through a comprehensive comparison of telescopic and meteor camera analyses, we can thus conclude that not only do the two sets of techniques agree on the properties of WJ1, they also both support that this object was the smallest asteroid compositionally characterized in space. We discussed a variety of `lessons learned' about to characterize Earth Impactors, hopefully future observers can use all of these techniques in combination to most effectively understand them.

\textbf{Acknowledgments}

The authors thank Myron Valenta for contributing the GMN data. Michael Mazur, Kasia Wisniewski, Yung Kipreos and Mariek Schmidt and her Brock University student search team are thanked for helping with the search in the weeks after the fall. Kathryn Turrentine is thanked for helping with aspects of the telescopic observations. Connell Miller and Araron Jaffe of the Northern Tornadoes Project, Western University, are thanked for their assistance with acquiring drone imagery and mapping of the main mass fall area. Finally, the authors thank Peter Jenniskens, Debora Rios, and Mariek Schmidt for helping with the search for smaller meteorites in May 2023. TK and NM were supported by the Mission Accessible Near-Earth Object Survey (MANOS), NASA grant number 80NSSC21K1328. DV was supported in part by NASA Cooperative Agreement 80NSSC21M0073 and by the Natural Sciences and Engineering Research Council of Canada. PJAM was supported by the Natural Sciences and Engineering Research Council of Canada. 
DF conducted this research at the Jet Propulsion Laboratory, California Institute of Technology, under a contract with the National Aeronautics and Space Administration (80NM0018D0004).


\vspace{5mm}
\facilities{LDT(LMI)}


\software{Wmpl}




%\bibliography{main}{}

\begin{thebibliography}{}
\expandafter\ifx\csname natexlab\endcsname\relax\def\natexlab#1{#1}\fi
\providecommand{\url}[1]{\href{#1}{#1}}
\providecommand{\dodoi}[1]{doi:~\href{http://doi.org/#1}{\nolinkurl{#1}}}
\providecommand{\doeprint}[1]{\href{http://ascl.net/#1}{\nolinkurl{http://ascl.net/#1}}}
\providecommand{\doarXiv}[1]{\href{https://arxiv.org/abs/#1}{\nolinkurl{https://arxiv.org/abs/#1}}}

\bibitem[{{Bida} {et~al.}(2014){Bida}, {Dunham}, {Massey}, \&
  {Roe}}]{2014SPIE.9147E..2NB}
{Bida}, T.~A., {Dunham}, E.~W., {Massey}, P., \& {Roe}, H.~G. 2014, in Society
  of Photo-Optical Instrumentation Engineers (SPIE) Conference Series, Vol.
  9147, Ground-based and Airborne Instrumentation for Astronomy V, ed. S.~K.
  {Ramsay}, I.~S. {McLean}, \& H.~{Takami}, 91472N, \dodoi{10.1117/12.2056872}

\bibitem[{{Bischoff} {et~al.}(2010){Bischoff}, {Horstmann}, {Pack},
  {Laubenstein}, \& {Haberer}}]{2010M&PS...45.1638B}
{Bischoff}, A., {Horstmann}, M., {Pack}, A., {Laubenstein}, M., \& {Haberer},
  S. 2010, \maps, 45, 1638, \dodoi{10.1111/j.1945-5100.2010.01108.x}

\bibitem[{Borovi{\v{c}}ka {et~al.}(2020)Borovi{\v{c}}ka, Spurn{\`y}, \&
  Shrben{\`y}}]{borovivcka2020two}
Borovi{\v{c}}ka, J., Spurn{\`y}, P., \& Shrben{\`y}, L. 2020, The Astronomical
  Journal, 160, 42

\bibitem[{Borovi{\v{c}}ka {et~al.}(2013)Borovi{\v{c}}ka, T{\'o}th, Igaz,
  Spurn{\`y}, Kalenda, Haloda, Svore{\v{n}}, Korno{\v{s}}, Silber, Brown,
  {et~al.}}]{borovivcka2013kovsice}
Borovi{\v{c}}ka, J., T{\'o}th, J., Igaz, A., {et~al.} 2013, Meteoritics \&
  Planetary Science, 48, 1757

\bibitem[{Borovi{\v{c}}ka {et~al.}(2015)Borovi{\v{c}}ka, Spurn{\`y},
  {\v{S}}egon, Andrei{\'c}, Kac, Korlevi{\'c}, Atanackov, Kladnik, Mucke, Vida,
  {et~al.}}]{borovivcka2015instrumentally}
Borovi{\v{c}}ka, J., Spurn{\`y}, P., {\v{S}}egon, D., {et~al.} 2015,
  Meteoritics \& Planetary Science, 50, 1244

\bibitem[{{Bowell} {et~al.}(1989){Bowell}, {Hapke}, {Domingue}, {Lumme},
  {Peltoniemi}, \& {Harris}}]{1989aste.conf..524B}
{Bowell}, E., {Hapke}, B., {Domingue}, D., {et~al.} 1989, in Asteroids II, ed.
  R.~P. {Binzel}, T.~{Gehrels}, \& M.~S. {Matthews}, 524--556

\bibitem[{{Bowen} {et~al.}(2023){Bowen}, {Reddy}, {De Florio}, {Kareta},
  {Pearson}, {Furfaro}, {Sharkey}, {McGraw}, {Cantillo}, {Sanchez}, \&
  {Battle}}]{2023PSJ.....4...52B}
{Bowen}, B., {Reddy}, V., {De Florio}, M., {et~al.} 2023, \psj, 4, 52,
  \dodoi{10.3847/PSJ/acb268}

\bibitem[{Brown {et~al.}(2010)Brown, Weryk, Kohut, Edwards, \&
  Krzeminski}]{brown2010development}
Brown, P., Weryk, R., Kohut, S., Edwards, W., \& Krzeminski, Z. 2010, WGN,
  Journal of the International Meteor Organization, 38, 25

\bibitem[{Brown {et~al.}(2011)Brown, McCausland, Fries, Silber, Edwards, Wong,
  Weryk, Fries, \& Krzeminski}]{brown2011fall}
Brown, P., McCausland, P., Fries, M., {et~al.} 2011, Meteoritics \& Planetary
  Science, 46, 339

\bibitem[{Brown {et~al.}(2023)Brown, McCausland, Hildebrand, Hanton, Eckart,
  Busemann, Krietsch, Maden, Welten, Caffee, {et~al.}}]{brown2023golden}
Brown, P.~G., McCausland, P., Hildebrand, A., {et~al.} 2023, Meteoritics \&
  Planetary Science, 58, 1773

\bibitem[{{Brunetto} {et~al.}(2015){Brunetto}, {Loeffler}, {Nesvorn{\'y}},
  {Sasaki}, \& {Strazzulla}}]{2015aste.book..597B}
{Brunetto}, R., {Loeffler}, M.~J., {Nesvorn{\'y}}, D., {Sasaki}, S., \&
  {Strazzulla}, G. 2015, in Asteroids IV, 597--616,
  \dodoi{10.2458/azu_uapress_9780816532131-ch031}

\bibitem[{{Bus} \& {Binzel}(2002)}]{2002Icar..158..146B}
{Bus}, S.~J., \& {Binzel}, R.~P. 2002, \icarus, 158, 146,
  \dodoi{10.1006/icar.2002.6856}

\bibitem[{{DellaGiustina} {et~al.}(2021){DellaGiustina}, {Kaplan}, {Simon},
  {Bottke}, {Avdellidou}, {Delbo}, {Ballouz}, {Golish}, {Walsh}, {Popescu},
  {Campins}, {Barucci}, {Poggiali}, {Daly}, {Le Corre}, {Hamilton}, {Porter},
  {Jawin}, {McCoy}, {Connolly}, {Garcia}, {Tatsumi}, {de Leon}, {Licandro},
  {Fornasier}, {Daly}, {Al Asad}, {Philpott}, {Seabrook}, {Barnouin}, {Clark},
  {Nolan}, {Howell}, {Binzel}, {Rizk}, {Reuter}, \&
  {Lauretta}}]{2021NatAs...5...31D}
{DellaGiustina}, D.~N., {Kaplan}, H.~H., {Simon}, A.~A., {et~al.} 2021, Nature
  Astronomy, 5, 31, \dodoi{10.1038/s41550-020-1195-z}

\bibitem[{{DeMeo} {et~al.}(2009){DeMeo}, {Binzel}, {Slivan}, \&
  {Bus}}]{2009Icar..202..160D}
{DeMeo}, F.~E., {Binzel}, R.~P., {Slivan}, S.~M., \& {Bus}, S.~J. 2009,
  \icarus, 202, 160, \dodoi{10.1016/j.icarus.2009.02.005}

\bibitem[{{DeMeo} \& {Carry}(2014)}]{2014Natur.505..629D}
{DeMeo}, F.~E., \& {Carry}, B. 2014, \nat, 505, 629,
  \dodoi{10.1038/nature12908}

\bibitem[{{Denneau} {et~al.}(2013){Denneau}, {Jedicke}, {Grav}, {Granvik},
  {Kubica}, {Milani}, {Vere{\v{s}}}, {Wainscoat}, {Chang}, {Pierfederici},
  {Kaiser}, {Chambers}, {Heasley}, {Magnier}, {Price}, {Myers}, {Kleyna},
  {Hsieh}, {Farnocchia}, {Waters}, {Sweeney}, {Green}, {Bolin}, {Burgett},
  {Morgan}, {Tonry}, {Hodapp}, {Chastel}, {Chesley}, {Fitzsimmons}, {Holman},
  {Spahr}, {Tholen}, {Williams}, {Abe}, {Armstrong}, {Bressi}, {Holmes},
  {Lister}, {McMillan}, {Micheli}, {Ryan}, {Ryan}, \&
  {Scotti}}]{2013PASP..125..357D}
{Denneau}, L., {Jedicke}, R., {Grav}, T., {et~al.} 2013, \pasp, 125, 357,
  \dodoi{10.1086/670337}

\bibitem[{Devillepoix {et~al.}(2020)Devillepoix, Cupak, Bland, Sansom, Towner,
  Howie, Hartig, Jansen-Sturgeon, Shober, Anderson,
  {et~al.}}]{devillepoix2020global}
Devillepoix, H., Cupak, M., Bland, P., {et~al.} 2020, Planetary and Space
  Science, 191, 105036

\bibitem[{{Devog{\`e}le} {et~al.}(2019){Devog{\`e}le}, {Moskovitz}, {Thirouin},
  {Gustaffson}, {Magnuson}, {Thomas}, {Willman}, {Christensen}, {Person},
  {Binzel}, {Polishook}, {DeMeo}, {Hinkle}, {Trilling}, {Mommert}, {Burt}, \&
  {Skiff}}]{2019AJ....158..196D}
{Devog{\`e}le}, M., {Moskovitz}, N., {Thirouin}, A., {et~al.} 2019, \aj, 158,
  196, \dodoi{10.3847/1538-3881/ab43dd}

\bibitem[{{Devog{\`e}le} {et~al.}(2024){Devog{\`e}le}, {Buzzi}, {Micheli},
  {Cano}, {Conversi}, {Jehin}, {Ferrais}, {Oca{\~n}a}, {F{\"o}hring}, {Drury},
  {Benkhaldoun}, \& {Jenniskens}}]{2024arXiv240404142D}
{Devog{\`e}le}, M., {Buzzi}, L., {Micheli}, M., {et~al.} 2024, arXiv e-prints,
  arXiv:2404.04142, \dodoi{10.48550/arXiv.2404.04142}

\bibitem[{Flynn {et~al.}(2018)Flynn, Consolmagno, Brown, \& Macke}]{Flynn2017}
Flynn, G., Consolmagno, G.~J., Brown, P., \& Macke, R.~J. 2018, Geochemistry,
  78, 269, \dodoi{10.1016/j.chemer.2017.04.002}

\bibitem[{Fries \& Fries(2010)}]{Fries2010}
Fries, M., \& Fries, J. 2010, Meteoritics \& Planetary Science, 45, 1476,
  \dodoi{10.1111/j.1945-5100.2010.01115.x}

\bibitem[{{Gaffey}(2010)}]{2010Icar..209..564G}
{Gaffey}, M.~J. 2010, \icarus, 209, 564, \dodoi{10.1016/j.icarus.2010.05.006}

\bibitem[{{Gaffey} {et~al.}(1993){Gaffey}, {Burbine}, \&
  {Binzel}}]{1993Metic..28..161G}
{Gaffey}, M.~J., {Burbine}, T.~H., \& {Binzel}, R.~P. 1993, Meteoritics, 28,
  161, \dodoi{10.1111/j.1945-5100.1993.tb00755.x}

\bibitem[{{Granvik} {et~al.}(2018){Granvik}, {Morbidelli}, {Jedicke}, {Bolin},
  {Bottke}, {Beshore}, {Vokrouhlick{\'y}}, {Nesvorn{\'y}}, \&
  {Michel}}]{2018Icar..312..181G}
{Granvik}, M., {Morbidelli}, A., {Jedicke}, R., {et~al.} 2018, \icarus, 312,
  181, \dodoi{10.1016/j.icarus.2018.04.018}

\bibitem[{Henych {et~al.}(2023)Henych, Borovi{\v{c}}ka, \&
  Spurn{\`y}}]{henych2023semi}
Henych, T., Borovi{\v{c}}ka, J., \& Spurn{\`y}, P. 2023, Astronomy and
  Astrophysics, 671, A23

\bibitem[{{Jenniskens} {et~al.}(2022){Jenniskens}, {Robertson}, {Goodrich},
  {Shaddad}, {Kudoda}, {Fioretti}, \& {Zolensky}}]{2022M&PS...57.1641J}
{Jenniskens}, P., {Robertson}, D., {Goodrich}, C.~A., {et~al.} 2022, \maps, 57,
  1641, \dodoi{10.1111/maps.13892}

\bibitem[{{Jenniskens} {et~al.}(2009){Jenniskens}, {Shaddad}, {Numan}, {Elsir},
  {Kudoda}, {Zolensky}, {Le}, {Robinson}, {Friedrich}, {Rumble}, {Steele},
  {Chesley}, {Fitzsimmons}, {Duddy}, {Hsieh}, {Ramsay}, {Brown}, {Edwards},
  {Tagliaferri}, {Boslough}, {Spalding}, {Dantowitz}, {Kozubal}, {Pravec},
  {Borovicka}, {Charvat}, {Vaubaillon}, {Kuiper}, {Albers}, {Bishop},
  {Mancinelli}, {Sandford}, {Milam}, {Nuevo}, \&
  {Worden}}]{2009Natur.458..485J}
{Jenniskens}, P., {Shaddad}, M.~H., {Numan}, D., {et~al.} 2009, \nat, 458, 485,
  \dodoi{10.1038/nature07920}

\bibitem[{Jenniskens {et~al.}(2018)Jenniskens, Albers, Tillier, Edgington,
  Longenbaugh, Goodman, Rudlosky, Hildebrand, Hanton, Ciceri,
  {et~al.}}]{jenniskens2018detection}
Jenniskens, P., Albers, J., Tillier, C.~E., {et~al.} 2018, Meteoritics \&
  Planetary Science, 53, 2445

\bibitem[{{Jenniskens} {et~al.}(2021){Jenniskens}, {Gabadirwe}, {Yin},
  {Proyer}, {Moses}, {Kohout}, {Franchi}, {Gibson}, {Kowalski}, {Christensen},
  {Gibbs}, {Heinze}, {Denneau}, {Farnocchia}, {Chodas}, {Gray}, {Micheli},
  {Moskovitz}, {Onken}, {Wolf}, {Devillepoix}, {Ye}, {Robertson}, {Brown},
  {Lyytinen}, {Moilanen}, {Albers}, {Cooper}, {Assink}, {Evers}, {Lahtinen},
  {Seitshiro}, {Laubenstein}, {Wantlo}, {Moleje}, {Maritinkole}, {Suhonen},
  {Zolensky}, {Ashwal}, {Hiroi}, {Sears}, {Sehlke}, {Maturilli}, {Sanborn},
  {Huyskens}, {Dey}, {Ziegler}, {Busemann}, {Riebe}, {Meier}, {Welten},
  {Caffee}, {Zhou}, {Li}, {Li}, {Liu}, {Tang}, {McLain}, {Dworkin}, {Glavin},
  {Schmitt-Kopplin}, {Sabbah}, {Joblin}, {Granvik}, {Mosarwa}, \&
  {Botepe}}]{2021M&PS...56..844J}
{Jenniskens}, P., {Gabadirwe}, M., {Yin}, Q.-Z., {et~al.} 2021, \maps, 56, 844,
  \dodoi{10.1111/maps.13653}

\bibitem[{Kikwaya {et~al.}(2010)Kikwaya, Weryk, Campbell-Brown, \&
  Brown}]{Kikwaya2010}
Kikwaya, J.~B., Weryk, R.~J., Campbell-Brown, M., \& Brown, P.~G. 2010, Monthly
  Notices of the Royal Astronomical Society, 398, 387,
  \dodoi{10.1111/j.1365-2966.2010.16294.x}

\bibitem[{{Lee}(1996)}]{1996Icar..124..181L}
{Lee}, P. 1996, \icarus, 124, 181, \dodoi{10.1006/icar.1996.0197}

\bibitem[{{Lu} \& {Jewitt}(2019)}]{2019AJ....158..220L}
{Lu}, X.-P., \& {Jewitt}, D. 2019, \aj, 158, 220,
  \dodoi{10.3847/1538-3881/ab4ce4}

\bibitem[{{Mainzer} {et~al.}(2011){Mainzer}, {Grav}, {Masiero}, {Hand},
  {Bauer}, {Tholen}, {McMillan}, {Spahr}, {Cutri}, {Wright}, {Watkins}, {Mo},
  \& {Maleszewski}}]{2011ApJ...741...90M}
{Mainzer}, A., {Grav}, T., {Masiero}, J., {et~al.} 2011, \apj, 741, 90,
  \dodoi{10.1088/0004-637X/741/2/90}

\bibitem[{McMullan {et~al.}(2023)McMullan, Vida, Devillepoix, Rowe, Daly, King,
  Cup{\'a}k, Howie, Sansom, Shober, {et~al.}}]{mcmullan2023winchcombe}
McMullan, S., Vida, D., Devillepoix, H.~A., {et~al.} 2023, Meteoritics \&
  Planetary Science

\bibitem[{McMullan {et~al.}(2024)McMullan, Vida, Devillepoix, Rowe, Daly, King,
  Cup{\'a}k, Howie, Sansom, Shober, {et~al.}}]{mcmullan2024winchcombe}
---. 2024, Meteoritics \& planetary science, 59, 927

\bibitem[{{Milliken}(2020)}]{2020pds..data...98M}
{Milliken}, R. 2020, {RELAB Spectral Library Bundle}, NASA Planetary Data
  System, urn:nasa:pds:relab::2.0, \dodoi{10.17189/1519032}

\bibitem[{{Mommert}(2017)}]{2017A&C....18...47M}
{Mommert}, M. 2017, Astronomy and Computing, 18, 47,
  \dodoi{10.1016/j.ascom.2016.11.002}

\bibitem[{{Moskovitz} {et~al.}(2022){Moskovitz}, {Wasserman}, {Burt},
  {Schottland}, {Bowell}, {Bailen}, \& {Granvik}}]{2022A&C....4100661M}
{Moskovitz}, N.~A., {Wasserman}, L., {Burt}, B., {et~al.} 2022, Astronomy and
  Computing, 41, 100661, \dodoi{10.1016/j.ascom.2022.100661}

\bibitem[{{Narita} {et~al.}(2020){Narita}, {Fukui}, {Yamamuro}, {Harbeck},
  {Bowman}, {Elphick}, {Nation}, {Armstrong}, {Han}, {Abe}, {Ikoma}, {Isogai},
  {Kawauchi}, {Kurita}, {Kusakabe}, {de Leon}, {Livingston}, {Mori},
  {Nishiumi}, {Tamura}, {Watanabe}, {Volgenau}, {Heinrich-Josties}, {Foale},
  {Daily}, {McCully}, {Kirby}, {Smith}, {Haworth}, {Conway},
  {Storrie-Lombardi}, {Rosing}, {Chatelain}, {Bachelet}, {Johnson}, \&
  {Rabus}}]{2020SPIE11447E..5KN}
{Narita}, N., {Fukui}, A., {Yamamuro}, T., {et~al.} 2020, in Society of
  Photo-Optical Instrumentation Engineers (SPIE) Conference Series, Vol. 11447,
  Ground-based and Airborne Instrumentation for Astronomy VIII, ed. C.~J.
  {Evans}, J.~J. {Bryant}, \& K.~{Motohara}, 114475K,
  \dodoi{10.1117/12.2559947}

\bibitem[{Ozerov {et~al.}(2024)Ozerov, Smith, Dotson, Longenbaugh, \&
  Morris}]{ozerov2024goes}
Ozerov, A., Smith, J.~C., Dotson, J.~L., Longenbaugh, R.~S., \& Morris, R.~L.
  2024, Icarus, 408, 115843

\bibitem[{Popova {et~al.}(2019)Popova, Borovicka, \&
  Campbell-Brown}]{popova2019modelling}
Popova, O., Borovicka, J., \& Campbell-Brown, M. 2019, Meteoroids. Sources of
  Meteors on Earth and Beyond, 9

\bibitem[{{Reddy} {et~al.}(2016){Reddy}, {Sanchez}, {Bottke}, {Thirouin},
  {Rivera-Valentin}, {Kelley}, {Ryan}, {Cloutis}, {Tegler}, {Ryan}, {Taylor},
  {Richardson}, {Moskovitz}, \& {Le Corre}}]{2016AJ....152..162R}
{Reddy}, V., {Sanchez}, J.~A., {Bottke}, W.~F., {et~al.} 2016, \aj, 152, 162,
  \dodoi{10.3847/0004-6256/152/6/162}

\bibitem[{Rumpf {et~al.}(2019)Rumpf, Longenbaugh, Henze, Chavez, \&
  Mathias}]{rumpf2019algorithmic}
Rumpf, C.~M., Longenbaugh, R.~S., Henze, C.~E., Chavez, J.~C., \& Mathias,
  D.~L. 2019, Sensors, 19, 1008

\bibitem[{{Sanchez} {et~al.}(2012){Sanchez}, {Reddy}, {Nathues}, {Cloutis},
  {Mann}, \& {Hiesinger}}]{2012Icar..220...36S}
{Sanchez}, J.~A., {Reddy}, V., {Nathues}, A., {et~al.} 2012, \icarus, 220, 36,
  \dodoi{10.1016/j.icarus.2012.04.008}

\bibitem[{Shaddad {et~al.}(2010)Shaddad, Jenniskens, Numan, Kudoda, Elsir,
  Riyad, Ali, Alameen, Alameen, Eid, {et~al.}}]{shaddad2010recovery}
Shaddad, M.~H., Jenniskens, P., Numan, D., {et~al.} 2010, Meteoritics \&
  Planetary Science, 45, 1557

\bibitem[{{Skamarock} {et~al.}(2019){Skamarock}, {Klemp}, {Dudhia}, {Gill},
  {Liu}, {Berner}, {Wang}, {Powers}, {Duda}, {Barker}, \&
  {Huang}}]{skamarock2019WRF4}
{Skamarock}, W.~C., {Klemp}, J.~B., {Dudhia}, J., {et~al.} 2019, A description
  of the advanced research WRF version 4, Tech. rep., NCAR Technical Note
  NCAR/TN-556+STR, \dodoi{10.5065/1dfh-6p97}

\bibitem[{Spurn{\`y} {et~al.}(2020)Spurn{\`y}, Borovi{\v{c}}ka, \&
  Shrben{\`y}}]{spurny2020vzvdar}
Spurn{\`y}, P., Borovi{\v{c}}ka, J., \& Shrben{\`y}, L. 2020, Meteoritics \&
  Planetary Science, 55, 376

\bibitem[{{Tatsumi} {et~al.}(2021){Tatsumi}, {Sugimoto}, {Riu}, {Sugita},
  {Nakamura}, {Hiroi}, {Morota}, {Popescu}, {Michikami}, {Kitazato},
  {Matsuoka}, {Kameda}, {Honda}, {Yamada}, {Sakatani}, {Kouyama}, {Yokota},
  {Honda}, {Suzuki}, {Cho}, {Ogawa}, {Hayakawa}, {Sawada}, {Yoshioka},
  {Pilorget}, {Ishida}, {Domingue}, {Hirata}, {Sasaki}, {de Le{\'o}n},
  {Barucci}, {Michel}, {Suemitsu}, {Saiki}, {Tanaka}, {Terui}, {Nakazawa},
  {Kikuchi}, {Yamaguchi}, {Ogawa}, {Ono}, {Mimasu}, {Yoshikawa}, {Takahashi},
  {Takei}, {Fujii}, {Yamamoto}, {Okada}, {Hirose}, {Hosoda}, {Mori}, {Shimada},
  {Soldini}, {Tsukizaki}, {Mizuno}, {Iwata}, {Yano}, {Ozaki}, {Abe}, {Ohtake},
  {Namiki}, {Tachibana}, {Arakawa}, {Ikeda}, {Ishiguro}, {Wada}, {Yabuta},
  {Takeuchi}, {Shimaki}, {Shirai}, {Hirata}, {Iijima}, {Tsuda}, {Watanabe}, \&
  {Yoshikawa}}]{2021NatAs...5...39T}
{Tatsumi}, E., {Sugimoto}, C., {Riu}, L., {et~al.} 2021, Nature Astronomy, 5,
  39, \dodoi{10.1038/s41550-020-1179-z}

\bibitem[{{Tonry} {et~al.}(2012){Tonry}, {Stubbs}, {Lykke}, {Doherty},
  {Shivvers}, {Burgett}, {Chambers}, {Hodapp}, {Kaiser}, {Kudritzki},
  {Magnier}, {Morgan}, {Price}, \& {Wainscoat}}]{2012ApJ...750...99T}
{Tonry}, J.~L., {Stubbs}, C.~W., {Lykke}, K.~R., {et~al.} 2012, \apj, 750, 99,
  \dodoi{10.1088/0004-637X/750/2/99}

\bibitem[{{Veach} {et~al.}(2022){Veach}, {Roming}, {Brody}, {Smith},
  {Killough}, {Persson}, {Pope}, {Peterson}, {Stange}, {Thibodeaux}, {Mathias},
  {Schwendeman}, {Thornton}, {Grubbs}, {Verastegui}, {Sutherland}, {Lechner},
  {Garc{\'\i}a-Vargas}, {Maldonado Medina}, {P{\'e}rez Calpena},
  {Sanchez-Blanco}, {Veredas}, {Robberto}, {van der Horst}, {Gelman}, {Smee},
  {Hope}, {Barkhouser}, \& {Koeppe}}]{2022SPIE12184E..68V}
{Veach}, T., {Roming}, P., {Brody}, A., {et~al.} 2022, in Society of
  Photo-Optical Instrumentation Engineers (SPIE) Conference Series, Vol. 12184,
  Ground-based and Airborne Instrumentation for Astronomy IX, ed. C.~J.
  {Evans}, J.~J. {Bryant}, \& K.~{Motohara}, 1218468,
  \dodoi{10.1117/12.2634061}

\bibitem[{{Vere{\v{s}}} {et~al.}(2012){Vere{\v{s}}}, {Jedicke}, {Denneau},
  {Wainscoat}, {Holman}, \& {Lin}}]{2012PASP..124.1197V}
{Vere{\v{s}}}, P., {Jedicke}, R., {Denneau}, L., {et~al.} 2012, \pasp, 124,
  1197, \dodoi{10.1086/668616}

\bibitem[{Vida {et~al.}(2024)Vida, Brown, Campbell-Brown, \&
  Egal}]{vida2024first}
Vida, D., Brown, P.~G., Campbell-Brown, M., \& Egal, A. 2024, Icarus, 408,
  115842

\bibitem[{Vida {et~al.}(2020)Vida, Gural, Brown, Campbell-Brown, \&
  Wiegert}]{vida2020}
Vida, D., Gural, P., Brown, P.~G., Campbell-Brown, M., \& Wiegert, P. 2020,
  Monthly Notices of the Royal Astronomical Society, 491, 2688,
  \dodoi{10.1093/mnras/stz3160}

\bibitem[{Vida {et~al.}(2021)Vida, {\v{S}}egon, Gural, Brown, McIntyre,
  Dijkema, Pavleti{\'c}, Kuki{\'c}, Mazur, Eschman, {et~al.}}]{vida2021global}
Vida, D., {\v{S}}egon, D., Gural, P.~S., {et~al.} 2021, Monthly Notices of the
  Royal Astronomical Society, 506, 5046

\bibitem[{Vida {et~al.}(2023)Vida, Brown, Devillepoix, Wiegert, Moser,
  Matlovi{\v{c}}, Herd, Hill, Sansom, Towner, {et~al.}}]{vida2023direct}
Vida, D., Brown, P.~G., Devillepoix, H.~A., {et~al.} 2023, Nature Astronomy, 7,
  318

\bibitem[{Voj{\'a}{\v{c}}ek {et~al.}(2022)Voj{\'a}{\v{c}}ek, Borovi{\v{c}}ka,
  \& Spurn{\`y}}]{vojavcek2022oxygen}
Voj{\'a}{\v{c}}ek, V., Borovi{\v{c}}ka, J., \& Spurn{\`y}, P. 2022, Astronomy
  and Astrophysics, 668, A102

\bibitem[{Weryk {et~al.}(2008)Weryk, Brown, Domokos, Edwards, Krzeminski,
  Nudds, \& Welch}]{weryk2008southern}
Weryk, R., Brown, P., Domokos, A., {et~al.} 2008, Advances in Meteoroid and
  Meteor Science, 241

\bibitem[{{Willmer}(2018)}]{2018ApJS..236...47W}
{Willmer}, C. N.~A. 2018, \apjs, 236, 47, \dodoi{10.3847/1538-4365/aabfdf}

\bibitem[{Wisniewski {et~al.}(2024)Wisniewski, Brown, Moser, \&
  Longenbaugh}]{Wisniewski2024}
Wisniewski, K., Brown, P., Moser, D., \& Longenbaugh, R. 2024, Icarus, 417,
  116118, \dodoi{10.1016/j.icarus.2024.116118}

\bibitem[{{Zellner} {et~al.}(1985){Zellner}, {Tholen}, \&
  {Tedesco}}]{1985Icar...61..355Z}
{Zellner}, B., {Tholen}, D.~J., \& {Tedesco}, E.~F. 1985, \icarus, 61, 355,
  \dodoi{10.1016/0019-1035(85)90133-2}

\bibitem[{{Zolensky}(2003)}]{2003ChEG...63..185Z}
{Zolensky}, M. 2003, Chemie der Erde / Geochemistry, 63, 185,
  \dodoi{10.1078/0009-2819-00038}

\bibitem[{{Zolensky} {et~al.}(2010){Zolensky}, {Herrin}, {Mikouchi}, {Ohsumi},
  {Friedrich}, {Steele}, {Rumble}, {Fries}, {Sandford}, {Milam}, {Hagiya},
  {Takeda}, {Satake}, {Kurihara}, {Colbert}, {Hanna}, {Maisano}, {Ketcham},
  {Goodrich}, {Le}, {Robinson}, {Martinez}, {Ross}, {Jenniskens}, \&
  {Shaddad}}]{2010M&PS...45.1618Z}
{Zolensky}, M., {Herrin}, J., {Mikouchi}, T., {et~al.} 2010, \maps, 45, 1618,
  \dodoi{10.1111/j.1945-5100.2010.01128.x}

\end{thebibliography}

\bibliographystyle{aasjournal}



\end{document}